\documentclass[10pt,journal,comsoc]{IEEEtran}
\usepackage{graphicx} 
\usepackage{graphics}
\usepackage{multirow}
\usepackage{textcomp}
\usepackage{array}

\usepackage{xcolor}
\usepackage{tabularx} 
\newcolumntype{Y}{>{\centering\arraybackslash}X} 
\usepackage{booktabs}

\usepackage{caption}
\usepackage{subcaption}

\makeatletter
\renewcommand\p@subfigure{\thefigure} 
\makeatother

\usepackage{tikz}
\usetikzlibrary{shapes.geometric, arrows.meta, positioning, fit, calc, backgrounds, shadows}

\usepackage{amsmath,amssymb,amsfonts}
\usepackage{bm}

\usepackage[algo2e, ruled, vlined, linesnumbered]{algorithm2e}

\usepackage{amsthm}
\theoremstyle{definition}

\SetKwProg{Fn}{Function}{}{end}

\SetKwInput{KwInit}{Init}
\SetAlgoSkip{}

\newcolumntype{M}[1]{>{\centering\arraybackslash}m{#1}}

\DeclareSymbolFont{symbolsC}{U}{txsyc}{m}{n}
\DeclareMathSymbol{\notniFromTxfonts}{\mathrel}{symbolsC}{61}

\title{Resilient Topology-Aware Coordination for Dynamic 3D UAV Networks under Node Failure}

\author{Chuan-Chi~Lai,~\IEEEmembership{Member,~IEEE}
    \IEEEcompsocitemizethanks{
        \IEEEcompsocthanksitem{This work was supported by the National Science and Technology Council (NSTC), Taiwan, under Grant No. NSTC 114-2221-E-194-062-, and was also partially supported by the Advanced Institute of Manufacturing with High-tech Innovations (AIM-HI) from the Featured Areas Research Center Program within the framework of the Higher Education Sprout Project by the Ministry of Education (MOE) in Taiwan. \emph{(Corresponding author: Chuan-Chi~Lai.)}}
        \IEEEcompsocthanksitem{C.-C. Lai is with the Department of Communications Engineering, National Chung Cheng University, Minxiong Township, Chiayi County 621301, Taiwan, and also with the Advanced Institute of Manufacturing with High-tech Innovations (AIM-HI), National Chung Cheng University, Minxiong Township, Chiayi County 621301, Taiwan (e-mail: chuanclai@ccu.edu.tw).}
    }
}

\IEEEtitleabstractindextext{%
\begin{abstract}
  Ensuring continuous service coverage under unexpected hardware failures is a fundamental challenge for 3D Aerial-Ground Integrated Networks. Although Multi-Agent Reinforcement Learning facilitates autonomous coordination, traditional architectures often lack resilience to sudden topology deformations. This paper proposes the Topology-Aware Graph MAPPO (TAG-MAPPO) framework to enhance system survivability through autonomous 3D spatial reconfiguration. Our framework integrates graph-based feature aggregation with a residual ego-state fusion mechanism to capture intricate inter-agent dependencies. To achieve structural robustness, we introduce a Random Observation Shuffling mechanism that fosters strong generalization to agent population fluctuations by breaking coordinate-index dependencies. Extensive simulations across heterogeneous environments, including high-speed mobility at 15 meters per second, demonstrate that TAG-MAPPO significantly outperforms Multi-Layer Perceptron baselines. Specifically, the framework reduces redundant handoffs by up to 50 percent while maintaining superior energy efficiency. Most notably, TAG-MAPPO exhibits exceptional self-healing capabilities, restoring over 90 percent of pre-failure coverage within 15 time steps. In dense urban scenarios, the framework achieves a post-failure fairness index surpassing its original four-UAV configuration by autonomously resolving service overlaps and interference. These findings confirm that topology-aware coordination is essential for resilient 6G aerial networks.
\end{abstract}

}
\begin{document}

\maketitle
\IEEEdisplaynontitleabstractindextext

%
\IEEEpeerreviewmaketitle

\section{Introduction}
\label{sec:intro}

\IEEEPARstart{T}{he} dawn of 6G networks signals a paradigm shift from terrestrial-dependent infrastructure to integrated, three-dimensional Non-Terrestrial Network (NTN) architectures~\cite{saad2020vision,giordani2021non}. Within this frontier, Unmanned Aerial Vehicles (UAVs) serving as Aerial Base Stations (ABSs) are a cornerstone solution due to their maneuverability, rapid deployment, and Line-of-Sight (LoS) connectivity~\cite{mozaffari2019tutorial,al2014modeling}. These networks are indispensable in mission-critical scenarios where terrestrial infrastructure is compromised, such as disaster relief or large-scale urban events~\cite{merwaday2015uav, zhao2019uav}. However, beyond non-stationary user distributions, ensuring system resilience against hardware malfunctions or catastrophic node failures remains a formidable obstacle for aerial orchestration.

To sustain coverage amidst volatility, Multi-Agent Proximal Policy Optimization (MAPPO) has been widely adopted~\cite{yu2022surprising, cui2020multi}. Nevertheless, a persistent limitation in RL-based UAV control is the structural rigidity of Multi-Layer Perceptron (MLP) architectures. Building upon our prior research, our previous work~\cite{lai2026spatiotemporalcontinuallearningmobile} addressed the plasticity-stability dilemma via Group-Decoupled MAPPO (G-MAPPO). While G-MAPPO mitigated gradient conflicts, its efficacy was constrained by the assumption of a quasi-static topology with a constant number of active nodes, leaving systems vulnerable to real-world disruptions. This paper extends that trajectory by pivoting toward topology-aware coordination to target resilience gaps left by static structural assumptions.

Practical aerial networks face dual-layer dynamics: continuous micro-temporal user displacement and discrete, catastrophic disruptions from UAV failures. MLP-based agents struggle with these ruptures due to fixed input dimensions and a lack of permutation invariance. When a neighbor fails, MLP architectures often suffer coordination collapse as they cannot dynamically re-weight spatial features, leading to service outages and failure to re-establish equilibrium.

To achieve self-healing communication, this paper proposes the Topology-Aware Graph Attention (TAG) framework. Graph-based learning is uniquely suited for wireless management due to its ability to generalize to varying node counts~\cite{dai2025asurvey2}. Instead of static feature vectors, we model the multi-UAV system as a dynamic graph~\cite{scarselli2008graph}. By incorporating a graph attention mechanism inspired by Graph Attention Networks (GAT)~\cite{velickovic2018graph, vaswani2017attention}, agents can dynamically assign importance weights to neighbors. This architecture, enhanced by a \textit{Random Observation Shuffling} (ROS) mechanism, enables the system to adapt to both user movements and abrupt topological changes.

The main contributions of this paper are summarized as follows: 
\begin{itemize} 
    \item \textbf{Topology-Aware Coordination Architecture:} We propose a MARL framework utilizing graph-based feature aggregation to capture relational structures. This enables spatial reasoning beyond simple observations, leading to stable coordination in interference-limited regimes. 
    \item \textbf{Signaling Stability and Resource Efficiency:} Integrating a handoff-penalized reward function with energy-efficient policy learning achieves a superior balance between coverage and signaling stability. Results show TAG-MAPPO reduces redundant handoffs by nearly 50 percent in sparse topologies while maintaining optimized energy efficiency. 
    \item \textbf{Autonomous Resilience via ROS:} We introduce a robust training paradigm incorporating ROS, regularizing the policy to be permutation-invariant and resilient to population fluctuations. By internalizing coordination logic under stochastic failures, the framework generalizes to unpredictable ruptures, restoring over 90 percent of service coverage following random node failures. 
    \item \textbf{Fairness Optimization via Structural Reconfiguration:} Our analysis reveals that in crowded urban environments, the framework strategically re-optimizes spatial deployment after failure to mitigate co-channel interference, effectively surpassing its pre-failure fairness performance.
\end{itemize}

The remaining sections of this paper are organized as follows. Section~\ref{sec:related_work} reviews related work on UAV deployment and MARL for UAV swarms. Sections~\ref{sec:system_model} and~\ref{sec:problem_formulation} respectively present the system model and problem formulation. Section~\ref{sec:proposed_framework} details the proposed TAG-MAPPO framework. Section~\ref{sec:simulation_results} discusses the simulation setup and performance evaluation. Finally, Section~\ref{sec:conclusion} concludes the paper.

\section{Related Work}
\label{sec:related_work}

This section reviews existing literature across three critical domains, including UAV coverage optimization, network resilience under node failure, and the integration of graph-based attention mechanisms within multi-agent reinforcement learning (MARL).

\subsection{Multi-UAV Control and Coverage Optimization}
The deployment of UAVs for wireless coverage has been studied via heuristic, mathematical, and learning-based approaches. Early research formulated the coverage problem as a facility location task, deriving optimal 3D locations for static users via convex optimization and circle packing theory~\cite{mozaffari2016efficient, al20173d}. To address energy constraints, trajectory optimization frameworks balanced throughput against propulsion energy consumption~\cite{zeng2017energy}. More recently, meta-heuristic algorithms have been applied to dynamic scenarios, such as collaborative differential evolution for path planning and energy-efficient graph algorithms for minimum-UAV coverage~\cite{Xu2025Evolving,Gong2024Energy}. However, these model-based methods typically assume global Channel State Information (CSI) availability and suffer from high computational latency, making them less suitable for real-time control in volatile environments.

With deep reinforcement learning (DRL) advancements, decentralized control strategies have emerged as a promising alternative. Since the introduction of Multi-Agent Deep Deterministic Policy Gradient (MADDPG)~\cite{lowe2017multi}, various DRL-based controllers have addressed adaptive 3D placement in 6G small cells~\cite{Hoang2025Adaptive} and joint trajectory design in space-air-ground integrated networks~\cite{10373024}. Beyond throughput, ensuring fairness has gained attention via adaptive deployment~\cite{Lai2023Adaptive} and utility-driven route adaptation for green cooperative networks~\cite{Aryendu2025AURA}. Despite these advances, most existing works rely on coordinate-based Multi-Layer Perceptrons (MLPs). This architecture struggles to capture the permutation-invariant nature of UAV swarms~\cite{zaheer2017deep}, often leading to coordination collapse when agent populations fluctuate.

\subsection{Network Resilience and Fault Tolerance} 
In mission-critical scenarios, system resilience, referring to the ability to recover from sudden disruptions, is paramount. Robustness in UAV search path planning has been enhanced using DRL for disaster environments~\cite{Liang2026Robustness}, while two-stage optimization has addressed collaborative post-disaster communications~\cite{Zheng2025UAV}. A recent study~\cite{Huang2026Achieving} addressed resilient topology configuration in 3D UAV networks using mixed-integer non-linear programming (MINLP) and recursive optimization to maintain connectivity. While providing theoretical bounds, such model-based approaches entail significant overhead during large-scale topological shifts.

To adapt to non-stationary environments, continual learning (CL) and distributed techniques have been adopted. A resilient topology optimization framework leveraging quantum annealing~\cite{Zhang2026Quantum} was introduced for rapid online switching, yet it relies on pre-defined candidate sets and specialized hardware. Building on foundational CL algorithms, a spatiotemporal-aware DRL framework~\cite{Chen2026Spatiotemporal} was introduced for cooperative coverage. In our prior work~\cite{lai2026spatiotemporalcontinuallearningmobile}, a spatiotemporal CL framework based on Group-Decoupled MAPPO (G-MAPPO) enabled UAVs to adapt to varying user distributions. However, current research predominantly focuses on macro-spatial variations. Micro-temporal resilience, involving immediate self-healing after node failure without pre-computed sets, remains under-explored. This paper addresses the gap by proposing a topology-aware controller for real-time fault recovery.

\subsection{Attention Mechanisms and Reconfigurable MARL} 
To overcome the limitations of fixed-input architectures, Graph Neural Networks (GNNs) and attention mechanisms have been integrated into MARL. Recent reviews emphasize that graph-based learning is uniquely suited for wireless resource management due to its inherent ability to generalize to dynamic network statuses and scale with varying numbers of nodes~\cite{dai2025asurvey2}. The effectiveness of using graph representations for resilient coordination is further supported by the topology sampling logic explored in~\cite{Zhang2026Quantum}. The Transformer architecture~\cite{vaswani2017attention} demonstrated the power of attention in sequence modeling, and it has been further shown that graph-based approaches possess strong relational inductive biases suitable for physical systems~\cite{battaglia2018relational}.

In the context of UAV mobility management, a hierarchical multi-agent DRL approach was proposed in~\cite{Meer2025Hierarchical} to handle dynamic clustering. Although hierarchical structures can be effective, they are often complex to train. Attention mechanisms offer a more scalable solution. Algorithms such as MAAC~\cite{iqbal2019actor} and MAGIC~\cite{niu2021multi} use attention to learn dynamic communication graphs. Nevertheless, few existing works explicitly integrate relative velocity into the attention mechanism for resilient trajectory control. By embedding velocity vectors and spatial relations, the proposed TAG-MAPPO framework enables agents to perceive the evolution of the network topology. This approach enhances stability in high-mobility environments and ensures that the remaining swarm can autonomously re-optimize coverage after a catastrophic node failure.

\section{System Model}
\label{sec:system_model}

\subsection{3D Aerial-Ground Network Architecture}
\label{subsec:architecture}

\subsubsection{Scenario Description}
\begin{figure}[!t]
    \centering
    \includegraphics[width=\columnwidth]{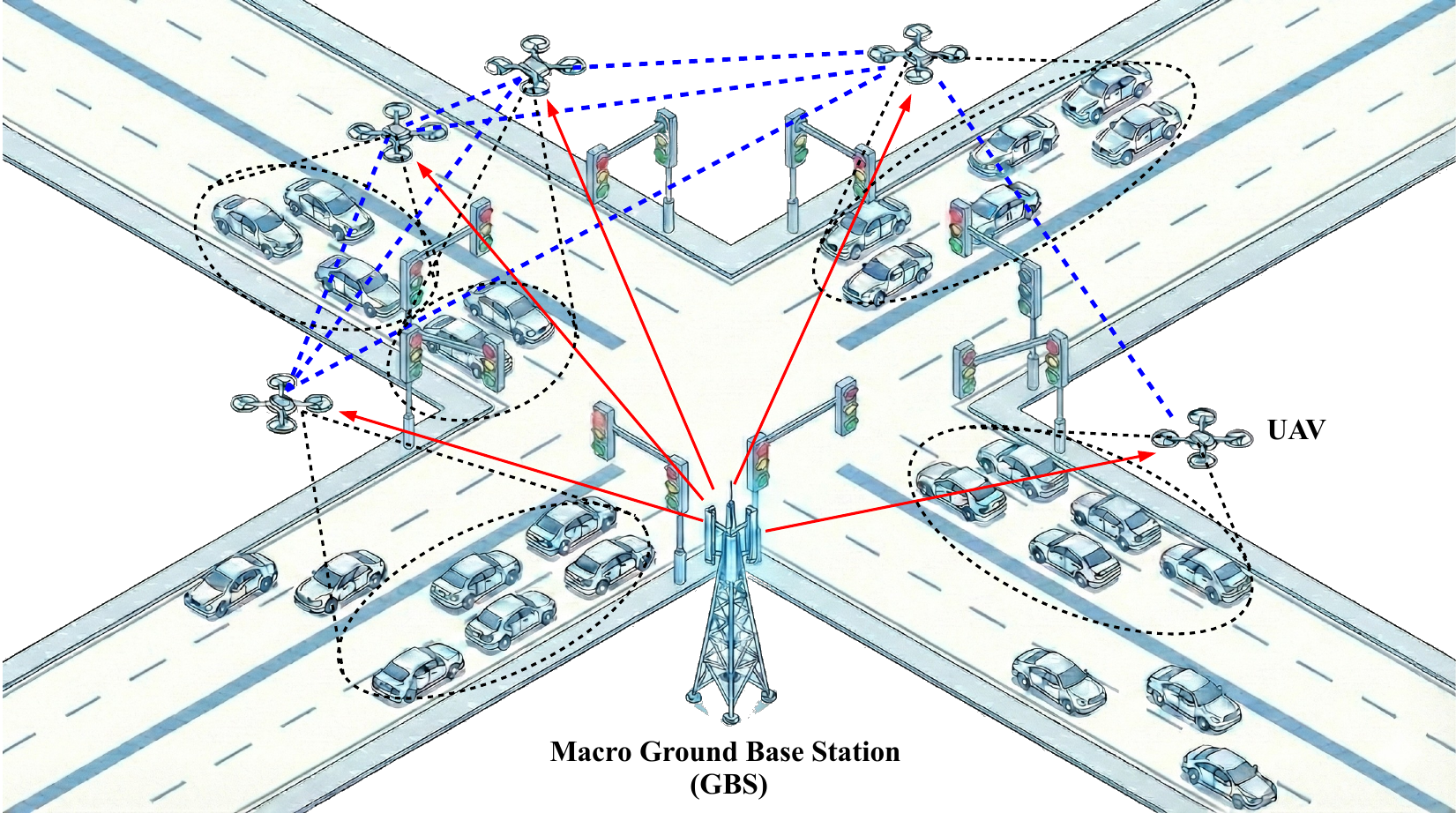} 
    \caption{Illustration of the dynamic 3D aerial-ground integrated network scenario. A fleet of UAVs functions as flying base stations to assist the Macro GBS in serving high-mobility vehicular platoons at a traffic intersection. The red solid arrows represent the high-capacity wireless backhaul links connected to the GBS (Data Plane), forming a star topology. The blue dotted lines denote the inter-UAV coordination links for exchanging local state information (Control Plane). The dashed ovals indicate the directional service footprints dynamically tracking the moving user clusters.}
    \label{fig:system_model}
\end{figure}

We consider a dynamic High-Mobility Edge Coverage scenario within 6G Non-Terrestrial Networks (NTN). As illustrated in Fig.~\ref{fig:system_model}, a fleet of Unmanned Aerial Vehicles (UAVs) operates as a flying small-cell tier to assist a macro Ground Base Station (GBS). This architecture is specifically designed for dynamic clusters such as vehicular platoons or emergency fleets to address severe spatiotemporal non-stationarity. The UAV swarm dynamically tracks user mobility to offload traffic and maintain reliable Line-of-Sight (LoS) connectivity, mitigating the inherent coverage limitations of terrestrial infrastructure.

\subsubsection{Physical Configuration and Survivability}
The system comprises $K_U$ rotary-wing UAVs and one stationary GBS, totaling $K = K_U + 1$ serving nodes. These nodes serve $M$ ground users distributed over a square target area $\mathcal{D} \subset \mathbb{R}^2$. Let $\mathcal{K}_U = \{1, \dots, K_U\}$ and $\mathcal{M} = \{1, \dots, M\}$ denote the sets of mobile UAVs and users, respectively. The GBS is situated at the center $\mathbf{p}_{\mathrm{GBS}} = [x_c, y_c, 0]^T$ to provide stable backhaul connectivity.

System resilience is characterized by a node survival indicator $\xi_k(t) \in \{0, 1\}$ for each UAV $k \in \mathcal{K}_U$. If UAV $k$ is operational at time $t$, $\xi_k(t) = 1$; otherwise, $\xi_k(t) = 0$ represents hardware failure or energy depletion. Since the GBS has a persistent power supply, we define $\xi_{\mathrm{GBS}}(t) = 1$ for all $t$.

The active subset of serving nodes at time $t$ is $\mathcal{V}(t) = \{k \in \mathcal{K}_U \mid \xi_k(t) = 1\} \cup \{\mathrm{GBS}\}$. An inactive UAV ($\xi_k(t)=0$) ceases all locomotion and transmission functions, removing it from the coordination process and interference environment. The state of each operational UAV $k \in \mathcal{V}(t) \setminus \{\mathrm{GBS}\}$ includes its 3D coordinates $\mathbf{p}_k(t) = [x_k(t), y_k(t), z_k(t)]^T$ and instantaneous velocity $\mathbf{v}_k(t)$, subject to a maximum flight speed $\|\mathbf{v}_k(t)\|\leq V_{\max}^{\text{UAV}}$ and a safe altitude corridor $[H_{\min}, H_{\max}]$.

\subsubsection{Inter-UAV Coordination and Control Plane}
To enable cooperative behavior, the UAV swarm and ground infrastructure are modeled as a dynamic graph $\mathcal{G}(t) = (\mathcal{V}(t), \mathcal{E}(t))$, where $\mathcal{V}(t)$ incorporates surviving UAV nodes and the stationary GBS. An undirected edge exists between operational UAVs if their distance is within the maximum communication range $R_{\mathrm{comm}}$. The GBS maintains persistent, long-range control-plane connectivity with all UAVs to facilitate global coordination and centralized value estimation.

Through these coordination links, agents exchange low-overhead state information, including instantaneous velocities and latent policy embeddings, within their local one-hop neighborhood. This decentralized exchange facilitates the proposed topology-aware attention mechanism, empowering the swarm to autonomously re-configure coordination weights whenever $\mathcal{G}(t)$ undergoes topological ruptures due to node failures.

\subsubsection{Wireless Backhaul (Data Plane)}
We assume a star topology for the data plane, where each surviving UAV $k \in \mathcal{V}(t) \setminus \{\text{GBS}\}$ maintains a high-capacity wireless backhaul link with the GBS at $\mathbf{p}_{\text{GBS}}$. The backhaul capacity is assumed to be sufficiently provisioned to support peak aggregate traffic. Consequently, system performance is limited by the dynamic fronthaul access links between UAVs and ground users rather than backhaul constraints.

\subsection{User Distribution and Mobility Models}
\label{subsec:mobility}
Ground user spatial dynamics involve initialization and temporal evolution phases contingent on the environmental scenario.

\subsubsection{Initial Spatial Distribution}
At $t=0$, user positions are initialized based on logical topology to simulate heterogeneous service requirements.
\begin{itemize}
  \item \textbf{Clustered (Urban/Suburban):} To represent hotspots, users are partitioned into $K_U+1$ clusters via K-Means. The cluster with the centroid closest to $\mathbf{p}_{\text{GBS}}$ is assigned to the terrestrial macro-cell, while the remaining $K_U$ clusters are designated as UAV target zones. Within each cluster, users follow a 2D Gaussian distribution $\mathcal{N}(\mathbf{c}_j, \sigma^2_c)$ around centroid $\mathbf{c}_j$.
  \item \textbf{Uniform (Rural):} In sparse scenarios, initial positions $\mathbf{q}_m(0)$ are uniformly distributed across the target area $\mathcal{D}$. Users connect to either the GBS or a UAV based on the Max-RSSI association policy.
\end{itemize}

\subsubsection{Mobility Dynamics}
We employ two mobility models to capture realistic movement patterns using specific kinematic notation.
\begin{itemize}
  \item \textbf{Reference Point Group Mobility (RPGM):} In high-density scenarios, user motion follows collective behavior. The position $\mathbf{q}_m(t)$ of user $m$ in cluster $j$ is $\mathbf{q}_m(t) = \mathbf{c}_j(t) + \mathbf{d}_{m,j}(t)$, where $\mathbf{c}_j(t)$ is the dynamic group center and $\mathbf{d}_{m,j}(t)$ represents a relative displacement vector simulating individual deviation.
  \item \textbf{Gauss-Markov (GM):} For independent users, we minimize abrupt velocity changes through the GM model. Let $\mathbf{v}_m(t)$ denote the velocity of user $m$. The update follows:
  \begin{equation}
    \mathbf{v}_m(t+1) = \alpha_{\text{GM}} \mathbf{v}_m(t) + (1-\alpha_{\text{GM}})\bar{\mathbf{v}}_m + \sqrt{1-\alpha_{\text{GM}}^2} \boldsymbol{\epsilon}_m(t),
  \end{equation}
  where $\alpha_{\text{GM}} \in [0, 1]$ is the memory parameter and $\bar{\mathbf{v}}_m$ is the asymptotic mean velocity. The term $\boldsymbol{\epsilon}_m(t)$ represents an independent Gaussian noise process.
\end{itemize}

\subsection{Channel Propagation Model}
\label{subsec:channel_model}
Communication performance is governed by the propagation characteristics of terrestrial and air-to-ground links.

\subsubsection{Terrestrial Channel (GBS-User)}
The path loss $L_{\text{GBS},m}(t)$ in dB between the GBS and user $m$ is modeled via log-normal shadowing:
\begin{equation}
  L_{\text{GBS},m}(t) = \text{PL}(d_0) + 10 \kappa \log_{10}\left(\frac{d_{\text{GBS},m}(t)}{d_0}\right) + \chi_{\sigma},
\end{equation}
where $\text{PL}(d_0)$ is the path loss at reference distance $d_0$, $\kappa$ is the path loss exponent, and $d_{\text{GBS},m}(t)$ is the 3D Euclidean distance. The term $\chi_{\sigma}$ accounts for shadow fading following a zero-mean Gaussian distribution with standard deviation $\sigma$.

\subsubsection{Air-to-Ground Channel (UAV-User)}
The link between an operational UAV $k \in \mathcal{V}(t)$ and user $m$ incorporates probabilistic Line-of-Sight (LoS) propagation. The LoS probability $P_{\mathrm{LoS},k,m}(t)$ depends on the elevation angle. The average path loss $\overline{\text{PL}}_{k,m}(t)$ is:
\begin{equation}
\begin{split}
  \overline{\text{PL}}_{k,m}(t) = & P_{\mathrm{LoS},k,m}(t)\eta_{\mathrm{LoS}} + (1 - P_{\mathrm{LoS},k,m}(t))\eta_{\mathrm{NLoS}} \\
  & + 20\log_{10}\left(\frac{4\pi f_c d_{k,m}(t)}{c}\right),
\end{split}
\end{equation}
where $\eta_{\mathrm{LoS}}$ and $\eta_{\mathrm{NLoS}}$ are excessive path loss coefficients for LoS and NLoS conditions, $f_c$ is the carrier frequency, and $c$ is the speed of light.

\subsubsection{Directional Antenna Gain}
Each UAV uses a directional antenna to suppress co-channel interference. Let $\phi_{k,m}(t)$ be the off-axis angle of user $m$ relative to the boresight of UAV $k$. The antenna gain $G_{\mathrm{tx}}(\phi_{k,m})$ follows a flat-top model:
\begin{equation}
  G_{\mathrm{tx}}(\phi_{k,m}) =
  \begin{cases}
    G_{\mathrm{main}}, & \text{if } |\phi_{k,m}(t)| \le \Theta_B \\
    G_{\mathrm{side}}, & \text{otherwise}
  \end{cases},
\end{equation}
where $G_{\mathrm{main}}$ and $G_{\mathrm{side}}$ are the main-lobe and side-lobe gains, and $2\Theta_B$ is the beamwidth. The effective channel gain is $h_{k,m}(t) = G_{\mathrm{tx}}(\phi_{k,m}) \cdot 10^{-\overline{\text{PL}}_{k,m}(t)/10}$.

\subsection{Interference Model and Data Rate}
\label{subsec:interference_model}
We assume a universal frequency reuse strategy where all operational UAVs in $\mathcal{V}(t)$ share the same frequency band $B$.

\subsubsection{Signal-to-Interference-plus-Noise Ratio (SINR)}
Under the Max-RSSI association policy where user $m$ connects to an active UAV $k \in \mathcal{V}(t) \setminus \{\text{GBS}\}$, the received SINR $\gamma_{k,m}(t)$ is:
\begin{equation}
  \gamma_{k,m}(t) = \frac{P_{\text{tx}} h_{k,m}(t)}{N_0 B + \sum_{j \in \mathcal{V}(t) \setminus {k, \text{GBS}}} P_{\text{tx}} h_{j,m}(t) + P_{\text{GBS}} h_{\text{GBS},m}(t)},
\end{equation}
where $P_{\text{GBS}} h_{\text{GBS},m}(t)$ represents co-channel interference from the terrestrial GBS. The summation in the denominator is restricted to the active set $\mathcal{V}(t) \setminus \{k, \text{GBS}\}$, ensuring that failed nodes do not contribute to the interference environment. Combined with persistent GBS interference, this formulation calculates aggregate interference only over surviving nodes.

\subsubsection{Achievable Data Rate}
Assuming equal bandwidth sharing among users associated with the same UAV, the downlink data rate $R_m(t)$ for user $m$ served by UAV $k$ is:
\begin{equation}
  R_m(t) = \frac{B}{N_k(t)} \log_2(1 + \gamma_{k,m}(t)),
\end{equation}
where $N_k(t)$ denotes the total load, representing the number of users currently served by UAV $k$.

\subsection{UAV Power Consumption Model}
We adopt an analytical power model for rotary-wing UAVs to evaluate network energy efficiency. Let $v_k(t) = \|\mathbf{v}_k(t)\|$ be the scalar speed of UAV $k$. The propulsion power $P_k^{\text{fly}}(t)$ required to maintain flight at speed $v_k(t)$ is formulated as~\cite{zeng2019energy}:
\begin{equation}\label{eq:propulsion_power}
\begin{split}
  P_k^{\text{fly}}(t) = & P_0 \left(1 + \frac{3 v_k(t)^2}{U_{\text{tip}}^2} \right) + \frac{1}{2} d_{\text{fuse}} \rho s A v_k(t)^3 \\
  & + P_i \left( \sqrt{1 + \frac{v_k(t)^4}{4 v_0^4}} - \frac{v_k(t)^2}{2 v_0^2} \right)^{1/2},
\end{split}
\end{equation}
where $P_0$ and $P_i$ are the blade profile and induced power in hovering, respectively. $U_{\text{tip}}$ is the rotor blade tip speed, and $v_0$ is the mean rotor induced velocity when hovering. The parameters $d_{\text{fuse}}$, $\rho$, $s$, and $A$ denote fuselage drag ratio, air density, rotor solidity, and rotor disc area. The three terms in \eqref{eq:propulsion_power} correspond to blade profile, induced, and parasitic power.

Total network power consumption $P_{\text{total}}(t)$ incorporates the energy cost of the UAV swarm and the ground infrastructure:
\begin{equation}\label{eq:total_power}
  P_{\text{total}}(t) = \sum_{k=1}^{K_U} \xi_k(t) (P_k^{\text{fly}}(t) + P_{\text{comm}}) + P_{\text{GBS}},
\end{equation}
where $K_U = 4$ and $K = K_U + 1$. $P_{\text{GBS}}$ denotes the constant power of the stationary GBS, including signal processing and backhaul circuitry. The survival indicator $\xi_k(t)$ ensures power is aggregated only over operational UAVs, as failed nodes do not draw propulsion or communication power.

\section{Problem Formulation}
\label{sec:problem_formulation}

\subsection{Network Utility Maximization}
The primary objective is to maintain service continuity and system stability amidst rapid user mobility and potential node failures. We define a composite utility function $U(t)$ derived from key performance indicators (KPIs) reflecting both Quality of Service (QoS) and Quality of Experience (QoE).

\subsubsection{System Capacity}
\label{subsubsec:throughput_sum}
To characterize the service capability of the aerial network, let $R_m(t)$ denote the instantaneous data rate of user $m$ at time $t$. The total system throughput is defined as the aggregate data rate of all ground users:
\begin{equation}
\label{eq:throughput_sum}
  R_{\text{sum}}(t) = \sum_{m=1}^{M} R_m(t).
\end{equation}
This metric reflects the capability of the swarm to manage co-channel interference and handle high-density traffic clusters. In the event of a node failure, a stable $R_{\text{sum}}(t)$ indicates that the surviving UAVs have successfully re-coordinated their spatial positions to compensate for the lost service capacity.

\subsubsection{Global Energy Efficiency}
\label{subsubsec:energy_efficiency}
To address the stringent power constraints of battery-limited UAVs, we maximize the Global Energy Efficiency (EE). This metric represents the bits delivered per Joule of energy consumed by the entire heterogeneous network $\mathcal{V}(t)$:
\begin{equation}
  \label{eq:energy_efficiency}
  E_{\text{ee}}(t) = \frac{R_{\text{sum}}(t)}{\sum_{k \in \mathcal{V}(t) \setminus {\text{GBS}}} (P_k^{\text{fly}}(t) + P_{\text{comm}}) + P_{\text{GBS}}},
\end{equation}
where $P_k^{\text{fly}}(t)$ is the propulsion power required for UAV $k$ to maintain flight at its instantaneous speed, while $P_{\text{GBS}}$ represents the constant power consumption of the stationary ground infrastructure. By evaluating $E_{\text{ee}}(t)$, we can assess whether the observed network resilience is achieved through energy-efficient maneuvers or via aggressive, high-speed propulsion that significantly reduces the endurance of the swarm.

\subsubsection{Coverage and QoS Satisfaction}
\label{subsubsec:coverage_qos}
We define the binary coverage status as $c_m(t) = \mathbb{I}(R_m(t) \ge R_{\text{th}})$, where $R_{\text{th}}$ is the minimum rate threshold. The network-wide coverage ratio is $C_{\text{cov}}(t) = \frac{1}{M} \sum_{m=1}^M c_m(t)$. Furthermore, we track the minimum user rate $R_{\min}(t) = \min_{m} R_m(t)$ to penalize worst-case performance, which is particularly critical during the reconfiguration phase following a node failure.

\subsubsection{Fairness Metrics}
\label{subsubsec:rate_fairness}
We employ Jain's Fairness Index to ensure equitable service among users. Rate Fairness $\mathcal{J}_{\text{R}}(t)$ ensures comparable data speeds to prevent specific users from suffering severe throughput degradation:
\begin{equation}\label{eq:rate_fairness}
  \mathcal{J}_{\text{R}}(t) = \frac{(\sum_{m=1}^{M} R_m(t))^2}{M \sum_{m=1}^{M} (R_m(t))^2 + \epsilon},
\end{equation}
where $\epsilon$ is a small constant for numerical stability.

\subsubsection{Service Continuity and Handoff Cost}
\label{subsubsec:handoff_cost}
Frequent handoffs (HO) induce significant signaling overhead and service interruptions. Let $u_m(t) \in \mathcal{V}(t)$ denote the association of user $m$ at time $t$. A handoff event occurs if $u_m(t) \neq u_m(t-1)$. We define the Handoff Cost as the total count of switching events:
\begin{equation}\label{eq:handoff_cost}
  C_{\text{ho}}(t) = \sum_{m=1}^{M} \mathbb{I}(u_m(t) \neq u_m(t-1)).
\end{equation}
Minimizing $C_{\text{ho}}(t)$ is crucial for maintaining robust connection tracking, especially when users must be reassigned to new nodes following the failure of their primary serving UAV.

\subsubsection{Composite Utility Function}
\label{subsubsec:utility_function}
The total utility $U(t)$ incorporates a stability penalty to discourage excessive handoffs:
\begin{equation}
  \label{eq:utility_function}
  U(t) = U_{\text{QoS}}(t) - \lambda_{\text{ho}} C_{\text{ho}}(t),
\end{equation}
where $U_{\text{QoS}}(t)$ is the weighted sum of normalized KPIs:
\begin{equation} \label{eq:utility_qos}
  \begin{split}
    U_{\text{QoS}}(t) = & \lambda_{\text{ee}} \tilde{E}_{\text{ee}}(t) + \lambda_{\text{jr}} \mathcal{J}_{\text{r}}(t) + \lambda_{\text{cov}} C_{\text{cov}}(t) \\
    & + \lambda_{\text{min}} \tilde{R}_{\text{min}}(t).
  \end{split}
\end{equation}
In this formulation, $\lambda{(\cdot)}$ represents weighting coefficients that balance aggressive throughput maximization against conservative stability maintenance. This hierarchical structure allows the proposed framework to prioritize mission endurance and service continuity during the rapid reconfiguration phases.

\subsection{Optimization Problem Statement and Feasibility Analysis}
We aim to optimize the joint 3D trajectories of all operational UAVs in $\mathcal{V}_U(t) = \mathcal{V}(t) \setminus \{\text{GBS}\}$ to maximize long-term time-averaged utility. The optimization problem is:
\begin{subequations}
\label{prob:p1}
\begin{align}
  (\textbf{P1}): \max_{\mathbf{P}} \quad & \lim_{T \to \infty} \frac{1}{T} \sum_{t=0}^{T} \mathbb{E}[U(t)] \label{eq:obj} \\
  \text{s.t.} \quad & \mathbf{p}_k(t+1) = \mathbf{p}_k(t) + \mathbf{v}_k(t) \Delta t, \quad \forall k \in \mathcal{V}_U(t) \label{eq:kinematic} \\
  & |\mathbf{v}_k(t)| \le V_{\max}^{\text{UAV}}, \quad \forall k \in \mathcal{V}_U(t) \label{eq:vel_constraint} \\
  & H_{\min} \le z_k(t) \le H_{\max}, \quad \forall k \in \mathcal{V}_U(t) \label{eq:alt_constraint} \\
  & \mathbf{p}_k(t) \in \mathcal{D}, \quad \forall k \in \mathcal{V}_U(t) \label{eq:bound_constraint} \\
  & |\mathbf{p}_i(t) - \mathbf{p}_j(t)| \ge d_{\text{safe}}, \quad \forall i, j \in \mathcal{V}(t), i \neq j \label{eq:safe_constraint}
\end{align}
\end{subequations}
The constraints in \eqref{prob:p1} define the physical boundaries under potential topological disruptions. Kinematic feasibility is governed by \eqref{eq:kinematic} and \eqref{eq:vel_constraint}. Flight corridor constraints \eqref{eq:alt_constraint} and \eqref{eq:bound_constraint} restrict UAV altitudes $z_k(t)$ and coordinates $\mathbf{p}_k(t)$ within compliant airspace. Collision avoidance in \eqref{eq:safe_constraint} ensures a minimum safety distance $d_{\text{safe}}$ between active nodes, including UAV-to-UAV and UAV-to-GBS separations. This is critical to prevent secondary failures during rapid reconfiguration when agents move aggressively to fill coverage gaps.

\textit{Computational Challenges and Feasibility:}
Problem (\textbf{P1}) presents challenges that render classical analytical frameworks impractical for real-time deployment. First, utility $U(t)$ involves intricate dependencies on coordinates, probabilistic channels, and dynamic interference, creating a non-convex landscape where gradient-based methods frequently converge to local optima. Second, the joint state-action space is subject to the curse of dimensionality, as complexity grows exponentially with users $M$ and surviving UAVs $\mathcal{V}_U(t)$. Third, high-mobility and node failures mean optimal trajectories derived for specific snapshots quickly become obsolete. Classical iterative methods, such as Successive Convex Approximation (SCA), require significant re-computation time upon topological ruptures, failing the millisecond-level responsiveness required for 6G applications.

Consequently, a low-latency, inference-based control paradigm is essential. By reformulating (\textbf{P1}) as a Deep Reinforcement Learning task, the computational burden shifts to offline training. During execution, the TAG-MAPPO framework provides near-instantaneous control through a single forward pass, ensuring real-time topology reconfiguration and rapid recovery.

\subsection{Reformulation as a Multi-Agent DEC-POMDP}
To address the intractability of (\textbf{P1}), we reformulate the control problem as a Multi-Agent Decentralized Partially Observable Markov Decision Process (DEC-POMDP) defined by the tuple $\langle \mathcal{K}, \mathcal{S}, \mathcal{A}, \mathcal{P}, \mathcal{R}, \Omega, \mathcal{O}, \gamma \rangle$.

\subsubsection{Global State Space $\mathcal{S}$}
The global state $s_t \in \mathcal{S}$ characterizes the complete network configuration at time $t$. To ensure the centralized critic evaluates node failures during training, the global state incorporates survival statuses. The state vector is:
\begin{equation}\label{eq:global_state}
  s_t = \left\{ \left\{ \mathbf{p}_k(t), \mathbf{v}_k(t), \xi_k(t) \right\}_{k \in \mathcal{K}_U}, \mathbf{p}_{\text{GBS}}, \left\{ \mathbf{q}_m(t), u_m(t) \right\}_{m \in \mathcal{M}} \right\},
\end{equation}
where $\xi_k(t)$ captures the operational status of each UAV $k \in \mathcal{K}_U$. By incorporating the GBS position $\mathbf{p}_{\text{GBS}}$ and dynamic user positions $\mathbf{q}_m(t)$, the critic estimates the value function based on spatial relationships between infrastructure, aerial nodes, and users. This representation allows the model to capture complex dependencies for centralized training and decentralized execution.

\subsubsection{Local Observation Space $\Omega$}
In decentralized execution, each operational UAV $k \in \mathcal{V}_U(t)$ makes decisions based on local observation $o_k(t)$. To achieve topology-invariant learning, absolute coordinates are transformed into a UAV-centric relative frame:
\begin{equation}\label{eq:local_observation}
  o_k(t) = \left[ \mathbf{x}_k^{\text{self}}, \mathbf{x}_k^{\text{GBS}}, \mathbf{X}_k^{\text{neigh}}, \mathbf{x}_k^{\text{user}} \right],
\end{equation}
where components capture specific environmental features:
\begin{itemize}
  \item \textbf{Self-State ($\mathbf{x}_k^{\text{self}}$):} Kinematic information of the agent, defined as $\mathbf{x}_k^{\text{self}} = [z_k(t), \mathbf{v}_k(t)]$. Altitude $z_k(t)$ and velocity $\mathbf{v}_k(t)$ are normalized by the safe corridor and $V_{\max}^{\text{UAV}}$.
  \item \textbf{GBS Anchor ($\mathbf{x}_k^{\text{GBS}}$):} Relative displacement to the fixed ground station, expressed as $\mathbf{x}_k^{\text{GBS}} = \mathbf{p}_{\text{GBS}} - \mathbf{p}_k(t)$. This allows the swarm to maintain a consistent coordinate origin after other nodes fail.
  \item \textbf{Neighborhood Topology ($\mathbf{X}_k^{\text{neigh}}$):} Relative features from neighboring UAVs within $R{\text{comm}}$. For each neighbor $j \in \mathcal{V}_U(t) \setminus \{k\}$, features include relative position $\Delta \mathbf{p}_{kj}(t) = \mathbf{p}_j(t) - \mathbf{p}_k(t)$ and velocity $\mathbf{v}_j(t)$. The set $\mathbf{X}_k^{\text{neigh}} = \{ [\Delta \mathbf{p}_{kj}(t), \mathbf{v}_j(t)] \}_{j \in \mathcal{V}_U(t) \setminus \{k\}}$ is processed by attention to re-weight coordination following failures.
  \item \textbf{User Distribution ($\mathbf{x}_k^{\text{user}}$):} Perceived as localized user densities or aggregated centroids within the footprint. They provide spatial cues for tracking high-mobility platoons and identifying service demand.
\end{itemize}

\subsubsection{Action Space $\mathcal{A}$}
The discrete action space $\mathcal{A}$ maps $a_k(t) \in \mathcal{A}$ to specific target velocity vectors. To simulate inertia and ensure smooth motion, the velocity $\mathbf{v}_k(t+1)$ is updated as:
\begin{equation}
  \mathbf{v}_k(t+1) = \beta \mathbf{v}_k(t) + (1-\beta) \mathbf{v}_{\text{target}}(a_k(t)) + \boldsymbol{\epsilon},
\end{equation}
where $\beta \in [0, 1)$ is the inertia coefficient and $\boldsymbol{\epsilon}$ is mechanical noise. Resulting trajectories remain continuous, and the target velocity norm is scaled to satisfy $\|\mathbf{v}_k(t+1)\| \leq V_{\max}^{\text{UAV}}$ for all operational UAVs in $\mathcal{V}_U(t)$.

\subsubsection{Cooperative Reward Function $\mathcal{R}$}
Agents share a common reward signal $r(t) = U(t)$ based on the utility in Section~\ref{subsubsec:utility_function}. The primary objective is to find an optimal joint policy $\boldsymbol{\pi}$ that maximizes the expected discounted return:
\begin{equation}
J(\boldsymbol{\pi}) = \mathbb{E} \left[ \sum_{t=0}^{\infty} \gamma^t r(t) \right],
\end{equation}
where $\gamma \in [0, 1)$ is the discount factor. Maximizing this return incentivizes surviving UAVs to autonomously re-configure their spatial topology to fill service gaps while maintaining stability. This collective reward mechanism ensures individual actions align with global performance objectives under node failure.

\section{Topology-Aware Graph-based Multi-Agent Reinforcement Learning (TAG-MAPPO)}
\label{sec:proposed_framework}

In this section, we propose the Topology-Aware Graph-based Multi-Agent Reinforcement Learning (TAG-MAPPO) framework. This framework is specifically engineered to surmount the challenges of high-mobility tracking and rapid topology deformation caused by both user movement and sudden node failures. To address the performance degradation typical of existing solutions in volatile environments, TAG-MAPPO adopts a Centralized Training with Decentralized Execution (CTDE) paradigm augmented by a specialized attention mechanism. Recent literature confirms that graph-based learning is uniquely suited for such dynamic wireless resource management tasks because of its inherent scalability and generalization capabilities~\cite{dai2025asurvey2}.

The design philosophy reconciles the need for real-time responsiveness on resource-constrained UAVs with the necessity for global coordination during training. Consequently, we introduce a lightweight Multi-Layer Perceptron (MLP) structure for decentralized actors to ensure low-latency execution. Conversely, a novel \textbf{Topology-Aware Graph Attention (TA-GAT)} mechanism is integrated into the centralized critic to tackle environmental non-stationarity. This architecture enables accurate value estimation from dynamic interaction graphs, which guides decentralized agents toward robust cooperative behaviors without imposing computational overhead during deployment.

\subsection{Overview of the TAG-MAPPO Architecture} 
\begin{figure*}[!t]
    \centering
    \includegraphics[width=\textwidth]{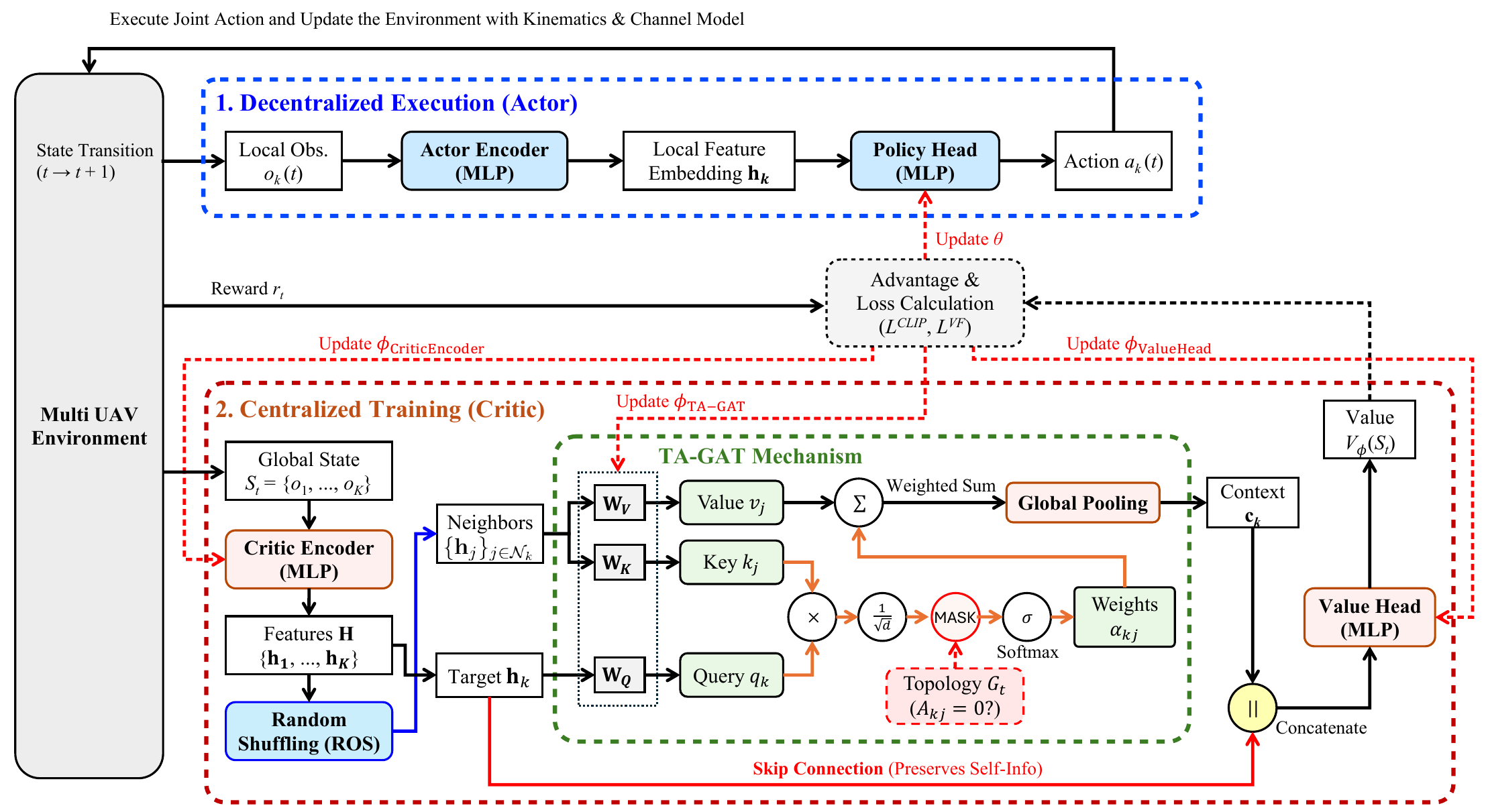} 
    \caption{The schematic architecture of the proposed TAG-MAPPO framework: (1) Decentralized Execution (Actor): Each UAV agent utilizes a lightweight, shared MLP encoder to map local observations $o_k(t)$ to actions, ensuring real-time responsiveness. (2) Centralized Training (Critic): The critic leverages a Topology-Aware Graph Attention (TA-GAT) mechanism to estimate state values $V_{\phi}(s_t)$. This module employs a decoupled dual-path strategy. A Random Observation Shuffling (ROS) operator is applied to the neighbor feature set $\{\mathbf{h}_j\}_{j \in \mathcal{N}_k(t)}$ to ensure permutation invariance. Simultaneously, a bold red skip connection preserves the ego-state information $\mathbf{h}_k$ to maintain operational consistency. This architecture enables the framework to perform precise relational reasoning and rapid spatial reconfiguration, which ensures robust coordination even under sudden node failures.}
    \label{fig:framework}
\end{figure*}

The overall architectural flow of the proposed TAG-MAPPO framework is illustrated in Fig.~\ref{fig:framework}. The processing pipeline is strictly structured to differentiate between the online execution phase and the offline centralized training phase.

The workflow consists of three distinct stages. In the first stage, each UAV operates independently using a shared policy network during deployment. By encoding raw local observations via a compact MLP, the system generates control actions directly. A critical advantage of this design is that the inference complexity remains constant irrespective of the global network size or user density, which ensures high scalability. In the second stage, the critic constructs a dynamic, ego-centric graph for each agent during the centralized training phase. Unlike standard methods that treat inputs as static vectors, the critic employs the TA-GAT module to aggregate features from the active neighbor set $\mathcal{V}(t) \setminus \{k\}$. This generates a permutation-invariant representation that captures high-order cooperative patterns such as load balancing and interference management. In the final stage, the value estimate generated by the TA-GAT critic serves as a baseline to compute the Generalized Advantage Estimation (GAE). This advantage signal acts as a teacher to guide the updates of the simplistic actor networks, which effectively distills the complex topological awareness of the critic into the lightweight actors via policy gradients.

\subsection{Dynamic Ego-Centric Graph Representation}
To formalize the time-varying interaction structure, the observation of the $k$-th UAV is modeled as a dynamic \textit{ego-centric graph} $\mathcal{G}_k(t) = (\mathcal{V}_k(t), \mathcal{E}_k(t))$. The node set $\mathcal{V}_k(t)$ incorporates the ego-UAV, the fixed GBS, and a dynamic set of neighboring UAV agents $\mathcal{N}_k(t)$ within communication range $R_{\mathrm{comm}}$. In high-mobility scenarios, the cardinality of the neighbor set $|\mathcal{N}_k(t)|$ is volatile. As UAVs and users adjust positions or when node failures occur, topological links are frequently established or broken. Standard MLPs struggle to process such variable-length inputs because they often require zero-padding, which introduces computational waste or numerical noise. In contrast, our graph-based formulation allows the critic to handle these topological fluctuations naturally. The attention mechanism can autonomously redistribute priority weights or ignore the absence of failed nodes, ensuring stable value estimation across fragmented network states.

\subsection{Feature Extraction and Embedding with Shuffling}
Before applying the attention mechanism, raw sensory data are transformed into high-level latent representations. Reflecting the CTDE architecture, distinct encoders are employed for decentralized actors and the centralized critic to cater to their specific functional roles.

For each decentralized actor, the local raw observation is mapped into a compact embedding vector via a lightweight MLP encoder. This encoder is shared across all UAV agents to facilitate parameter efficiency and cooperative behavior. The resulting local embedding encapsulates the immediate kinematic state required for reactive control and serves as the primary input for the policy head.

The centralized critic employs a sophisticated encoding scheme to handle global state information. To ensure the learned policy remains robust against agent indexing, we introduce a Random Observation Shuffling (ROS) mechanism during training. Before entering the TA-GAT layer, the neighbor feature set is processed by a stochastic permutation operator to achieve Permutation Invariance:
\begin{equation} 
  \tilde{\mathbf{X}}_k^{\text{neigh}}(t) = \text{StochasticPermutation}\left( \{ \mathbf{h}_j(t) \}_{j \in \mathcal{N}_k(t)} \right). 
\end{equation} 
This ensures the critic learns to prioritize neighbors based on physical attributes rather than their position in the input tensor. Crucially, ROS is applied only to the neighboring agent set $\{\mathbf{h}_j(t)\}_{j \in \mathcal{N}_k(t)}$, while the ego-state feature $\mathbf{h}_k(t)$ remains fixed as an anchor for the attention query. This decoupled shuffling strategy forces the attention mechanism to learn relative topological relationships between the target UAV and its surrounding swarm. By treating the neighborhood as an unordered collection of relational entities, the policy becomes resilient to dynamic changes in the agent population.

This design directly contributes to the system's self-healing capabilities. When a neighboring node fails, remaining agents maintain stable coordination because the learned feature aggregation is independent of specific neighbor indices. By projecting these diverse and permuted inputs into a unified feature space, the encoders enable the TA-GAT layer to perform effective relational reasoning under varying topological densities and volatile operational conditions.

\subsection{Topology-Aware Graph Attention Mechanism (TA-GAT)}
A fundamental challenge in applying Graph Neural Networks to UAV control is the potential dilution of the agent's own state information during neighbor feature aggregation. To address topology deformation, we introduce a Dual-Path Aggregation strategy within the TA-GAT layer.

\subsubsection{Attention-based Neighbor Aggregation}
The first path focuses on understanding the environment. The relevance of each neighbor $j$ to the ego-agent $k$ is computed using an attention mechanism. Features are projected into Query, Key, and Value spaces using learnable weight matrices $\mathbf{W}_Q, \mathbf{W}_K$, and $\mathbf{W}_V$. The attention score $e_{kj}(t)$, which represents the importance of neighbor $j$, is computed using the Scaled Dot-Product Attention:
\begin{equation}\label{eq:attention_score}
  e_{kj}(t) = \frac{(\mathbf{W}_Q \mathbf{h}_k(t))^T (\mathbf{W}_K \mathbf{h}_j(t))}{\sqrt{d_{\text{attn}}}}.
\end{equation}
In this expression, $d_{\text{attn}}$ denotes the dimensionality of the query and key vectors. The scaling factor $1/\sqrt{d_{\text{attn}}}$ is employed to counteract the effect of large dot-product values, which would otherwise push the softmax function into regions with extremely small gradients.

To handle the variable number of neighbors, a topology mask is applied to the attention scores. These scores are then normalized via the softmax function to obtain attention weights $\alpha_{kj}(t)$. These weights are subsequently used to aggregate neighbor features into a \textit{Neighbor Context Vector} $\mathbf{c}_{\text{neigh}}(t)$:
\begin{equation}\label{eq:neighbor_aggregation}
  \mathbf{c}_{\text{neigh}}(t) = \sum_{j \in \mathcal{V}(t) \setminus \{k\}} \alpha_{kj}(t) (\mathbf{W}_V \mathbf{h}_j(t)).
\end{equation}
This vector summarizes the collective influence of the neighborhood, which is weighted by kinematic relevance such as relative velocity and distance to the ego-agent.

\subsubsection{Dual-Path Fusion and Skip Connection}
To prevent the dilution of the agent's own kinematic features, we employ a skip connection architecture. The processed ego-feature is concatenated directly with the aggregated neighbor context. The final topology-aware embedding $\mathbf{g}_k(t)$ is defined as:
\begin{equation}\label{eq:dual_path_fusion}
  \mathbf{g}_k(t) = \text{MLP}_{\text{head}} \left( [ \mathbf{W}_{\text{self}} \mathbf{h}_k(t) \mathbin\Vert \mathbf{c}_{\text{neigh}}(t) ] \right),
\end{equation}
where $\mathbin\Vert$ denotes the concatenation operation. This structure ensures that the critic maintains a sharp awareness of the agent's own capabilities while fully accounting for the topological influence of its neighbors. This fused feature is passed to the value head to estimate the state value $V_{\phi}(s_t)$.

\subsection{Multi-Agent Actor-Critic Learning}
The proposed TA-GAT mechanisms are integrated into the MAPPO framework to enable stable learning in volatile environments. Each UAV agent $k$ optimizes its local policy $\pi_{\theta}$ to maximize the clipped PPO objective:
\begin{equation}\label{eq:ppo_clip}
  L(\theta) = \mathbb{E}_{k, t} \left[ \min( \rho_k(t) A_k(t), \text{clip}(\rho_k(t), 1-\epsilon, 1+\epsilon) A_k(t) ) \right],
\end{equation}
where $\rho_k(t)$ is the probability ratio between the new and old policies, and $A_k(t)$ is the advantage calculated via GAE based on the TA-GAT value estimate.

To enhance training stability against potential value outliers caused by rapid topology changes or sudden node failures, we employ the \textbf{Huber Loss} for the critic update:
\begin{equation}\label{eq:huber_loss}
  \mathcal{L}(\phi) = \begin{cases} \frac{1}{2} (y_t - V_{\phi}(s_t))^2 & \text{for } |y_t - V_{\phi}(s_t)| \leq \delta, \\ \delta (|y_t - V_{\phi}(s_t)| - \frac{1}{2}\delta) & \text{otherwise}, \end{cases}
\end{equation}
where $y_t$ denotes the temporal difference target and $\phi$ represents the learnable parameters of the TA-GAT critic. The hyperparameter $\delta$ serves as a threshold that determines the transition point from quadratic to linear loss. This formulation provides enhanced robustness by behaving quadratically for small estimation errors while maintaining a constant gradient for large errors. Such a property ensures that the gradient updates remain stable even during drastic topology shifts or reward spikes triggered by unexpected node failures.

\subsection{TAG-MAPPO Training Algorithm}
The complete training procedure for the TAG-MAPPO framework is formalized in Algorithm \ref{alg:tag_marl}. The optimization pipeline consists of decentralized data collection and centralized parameter updates.

\begin{algorithm2e}[!t]
\caption{Training Procedure for TAG-MAPPO with ROS}
\label{alg:tag_marl}
\SetAlgoLined
\SetKwInput{KwInit}{Initialize}
\SetKwInput{KwParam}{Hyperparameters}
\KwInit{Actor network $\pi_\theta$, Critic network $V_\phi$, Replay buffer $\mathcal{D}$, Optimizers.}
\KwParam{Learning rate $\eta$, Entropy coefficient $\beta$, Clip ratio $\epsilon$, Huber threshold $\delta$.}
\BlankLine
\For{episode $m = 1$ \KwTo $M$}{
    Reset environment and obtain initial observations $\mathbf{o}(0)$ for all active UAVs in $\mathcal{V}(0)$\;
    $\mathcal{D} \leftarrow \emptyset$\;
    \tcp{Phase 1: Decentralized Data Collection}
    \For{step $t = 0$ \KwTo $T-1$}{
        \For{each operational agent $k \in \mathcal{V}(t)$}{
            Extract local features: $\mathbf{h}_k(t) \leftarrow \text{Enc}_{\text{actor}}(o_k(t))$\;
            Sample action from policy: $a_k(t) \sim \pi_{\theta}(\cdot | \mathbf{h}_k(t))$\;
        }
        Execute joint actions $\mathbf{a}(t)$, observe reward $r(t)$ and next observations $\mathbf{o}(t+1)$\;
        \textbf{Identify topological changes}: Update surviving set $\mathcal{V}(t+1)$ if node failure occurs\;
        Store transition $(\mathcal{V}(t), \mathbf{o}(t), \mathbf{a}(t), r(t), \mathbf{o}(t+1))$ in $\mathcal{D}$\;
    }
    \tcp{Phase 2: Centralized Parameter Update}
    Compute value estimates $V_\phi(s_t)$ and GAE advantages $\hat{A}_t$ using the TA-GAT critic based on captured $\mathcal{V}(t)$\;
    \For{epoch $j = 1$ \KwTo $K_{\text{epochs}}$}{
        Generate random minibatches from $\mathcal{D}$\;
        \For{each minibatch $\mathcal{B} \subset \mathcal{D}$}{
            \For{each agent $k$ in minibatch $\mathcal{B}$}{
                \textbf{Apply Random Observation Shuffling (ROS)}: $\tilde{\mathcal{N}}_k(t) \leftarrow \text{permute}(\mathcal{N}_k(t))$\;
                Construct ego-centric graph $\mathcal{G}_k(t)$ and compute attention weights $\alpha_{kj}(t)$\;
                Aggregate context $\mathbf{c}_{\text{neigh}}(t)$ and perform dual-path fusion via \eqref{eq:dual_path_fusion}\;
            }
            Update Critic $\phi$ by minimizing Huber loss $\mathcal{L}(\phi)$ based on TD-error\;
            Update Actor $\theta$ by maximizing PPO-Clip objective $L(\theta)$ with entropy bonus $\beta$\;
        }
    }
    Anneal entropy coefficient $\beta$ and learning rates $\eta$ linearly\;
}
\end{algorithm2e}

In the first phase, distributed UAV agents interact with the environment using lightweight actor networks. At each time step, agents rely on local observations to sample actions, ensuring inference latency remains constant regardless of network size. Upon node failures, the surviving swarm $\mathcal{V}(t)$ must immediately adapt based on remaining local observations. Transition data, including the instantaneous topology state $\mathcal{V}(t)$, is stored in the replay buffer to provide the critic with necessary context for post-failure value estimation.

In the second phase, the system executes centralized updates. Before optimization, Generalized Advantage Estimation (GAE) is computed to balance the bias-variance trade-off. The TA-GAT critic plays a pivotal role by reconstructing dynamic ego-centric graphs for each minibatch sample. Specifically, the ROS mechanism is applied to the neighbor feature set before aggregation. By stochastically reordering neighbors, the critic learns relational features invariant to agent indexing. Through dual-path aggregation, the critic predicts state values to compute the Huber-based value loss. This mechanism ensures that lightweight actors become topology-aware by distilling complex relational knowledge from the critic into the simplistic policy networks.

\subsection{Computational Complexity Analysis}
During the online execution phase, computational complexity is dominated by matrix-vector multiplications within the MLP-based actors, scaling as $\mathcal{O}(L \cdot D^2)$ where $L$ is the number of layers and $D$ is the hidden dimension. This complexity is independent of the number of dynamic neighbors, satisfying the real-time constraints required for 6G mission-critical tasks.

During the offline training phase, complexity is primarily driven by the attention mechanism within the centralized critic. For each agent, the TA-GAT module involves linear projections and scaled dot-product computations. For an ego-centric graph, this scales as $\mathcal{O}((|\mathcal{V}(t)| + |\mathcal{N}_k(t)|) \cdot D)$, where $|\mathcal{N}_k(t)|$ is the number of active neighbors. While the attention operation involves more intensive processing than simple MLPs, its execution is restricted to the centralized training server with substantial resources. Consequently, TAG-MAPPO achieves an optimal trade-off by utilizing topology-aware reasoning offline to distill robust policies into efficient, lightweight actors for deployment.

\section{Simulation Results and Analysis}
\label{sec:simulation_results}

In this section, we conduct extensive simulations to evaluate the performance and resilience of the proposed TAG-MAPPO framework. We outline the simulation setup and analyze experimental results across various environmental scenarios, focusing on system recovery following node failures.

\subsection{Simulation Setup and Parameters}
The simulation environment is developed using a customized Python-based multi-UAV simulator. The target area is a $1000 \times 1000 \text{ m}^2$ square region served by a heterogeneous network consisting of $K_{U}=4$ mobile UAVs and one stationary ground base station (GBS), totaling $K=5$ serving nodes for $M=140$ users. The UAV flight altitude corridor is $[H_{\min}, H_{\max}] = [80, 120] \text{ m}$. Maximum speeds for UAVs and independent ground users are capped at $V_{\max}^{\text{UAV}} = 30 \text{ m/s}$ and $V_{\max}^{\text{user}} = 15 \text{ m/s}$, respectively. In clustered scenarios, the group velocity for vehicular platoons is maintained at $10 \text{ m/s}$.

To ensure balanced learning across objectives, individual reward components in \eqref{eq:utility_qos} are normalized to $[0, 1]$. The weighting coefficients are $(\lambda_{\text{cov}}, \lambda_{\text{ee}}, \lambda_{\text{jr}}, \lambda_{\text{min}}) = (3.5, 0.1, 0.5, 0.5)$, prioritizing global coverage while maintaining service quality for worst-case users. Additionally, the handoff penalty $\lambda_{\text{ho}}$ is tuned to $0.1$ to suppress redundant signaling and maintain stability during reconfiguration.

To evaluate resilience, a catastrophic node failure is triggered at $t=100$ by setting $\xi_k(t) = 0$ for a randomly selected UAV node (excluding the GBS), forcing the remaining swarm to reconfigure autonomously. Consolidated parameters are detailed in Table \ref{tab:parameters}.

\subsection{Benchmark Scenarios and Channel Characterization}
To evaluate generalization and robustness, we define three simulation scenarios based on 3GPP propagation models. The specific LoS/NLoS characteristics for each environment are detailed as follows:
\begin{itemize}
  \item \textbf{Crowded Urban:} This interference-limited regime features high user density with channel parameters $(a=12.08, b=0.11)$ and a significant NLoS path loss exponent ($\eta_{\text{nlos}}=23.0$). The primary challenge is managing complex co-channel interference within user clusters modeled via RPGM mobility.
  \item \textbf{Suburban:} This scenario involves moderate user dispersion and balanced LoS probability ($a=9.61, b=0.16$). It serves as the baseline for evaluating the optimal resilience of the UAV swarm, where the topology allows for effective spatial reuse.
  \item \textbf{Rural Area:} This noise-limited environment features sparse, Gauss-Markov independent mobility with high LoS probability ($a=4.88, b=0.43$). Lower blockage frequency simplifies link optimization but tests the over-smoothing limits of the graph coordination mechanism.
\end{itemize}

\begin{table}[!t]
\centering
\caption{Consolidated Simulation Parameters and Hyperparameter Settings}
\label{tab:parameters}
\begin{tabular}{lcl}
\hline
\textbf{Parameter (Physical)} & \textbf{Symbol} & \textbf{Value} \\ \hline
Number of UAVs & $K_U$ & 4 \\
Ground Base Station (GBS) & - & 1 \\
UAV Max Speed & $V_{\max}^{\text{UAV}}$ & 30 m/s \\
User Max Speed & $V_{\max}^{\text{user}}$ & 15 m/s \\
Carrier Frequency & $f_c$ & 2.0 GHz \\
System Bandwidth & $B$ & 20 MHz \\
Transmission Power & $P_{\text{tx}}$ & 23 dBm \\
Noise PSD & $N_0$ & $-174$ dBm/Hz \\
Altitude Range & $[H_{\min}, H_{\max}]$ & $[80, 120]$ m \\
GBS Fixed Power & $P_{\text{GBS}}$ & 20 W \\
UAV Comm. Power & $P_{\text{comm}}$ & 5 W \\
Target Area & $\mathcal{D}$ & $1 \times 1 \text{ km}^2$ \\
Safety Distance & $d_{\text{safe}}$ & 10 m \\
Number of Users & $M$ & 140 \\ \hline
\textbf{Hyperparameter (DRL)} & \textbf{Symbol} & \textbf{Value} \\ \hline
Actor Learning Rate & $\alpha_a$ & $1 \times 10^{-4}$ \\
Critic Learning Rate & $\alpha_c$ & $5 \times 10^{-4}$ \\
Discount Factor & $\gamma$ & 0.99 \\
GAE Parameter & $\tau$ & 0.95 \\
PPO Epochs & $E_{\text{ppo}}$ & 10 \\
Batch Size & $B_{\text{size}}$ & 256 \\
Huber Loss Delta & $\delta$ & 2.0 \\
Entropy Coeff. & $\beta_{\text{ent}}$ & $0.10 \to 0.01$ \\ 
Total number of episodes & $M$ & 5000 \\ 
Episode index & $ep$ & 1, 2, ..., 5000 \\ \hline
\textbf{Reward Weights} & \textbf{Symbol} & \textbf{Value} \\ \hline
Coverage & $\lambda_{\text{cov}}$ & 3.5 \\
Energy Efficiency & $\lambda_{\text{ee}}$ & 0.1 \\
Rate Fairness & $\lambda_{\text{jr}}$ & 0.5 \\
Minimum Rate & $\lambda_{\text{min}}$ & 0.5 \\
Handoff Penalty & $\lambda_{\text{ho}}$ & 0.1 \\ \hline
\end{tabular}
\end{table}

\subsection{Performance Metrics}
To quantify the system's performance and its adaptive behavior under disruptions, we evaluate the following four key metrics, leveraging the formal definitions provided in Section \ref{sec:problem_formulation}:

\begin{itemize}
  \item \textbf{Network Resilience ($C_{\text{cov}}$ \& $\mathcal{J}_{\text{R}}$):} We evaluate the post-failure recovery of the \textit{Coverage Ratio} and \textit{Jain's Fairness Index} as defined in Sections \ref{subsubsec:coverage_qos} and \ref{subsubsec:rate_fairness}, respectively. These serve as the primary indicators of the swarm's ``survivability'' and its autonomous ability to re-equilibrate service distribution after a loss of hardware resources.
  \item \textbf{Topological Stability ($C_{\text{ho}}$):} This represents the total handoff count per episode. As defined in Section \ref{subsubsec:handoff_cost}, this metric is critical for assessing whether the controller maintains service through stable spatial positioning or via erratic user-UAV re-associations that would trigger signaling storms in real-world 6G deployments.
  \item \textbf{Operational Efficiency ($E_{\text{ee}}$):} Defined in Section \ref{subsubsec:energy_efficiency}, this global energy efficiency metric reflects the ratio of system throughput to the total power consumption of the \textbf{active} UAV set $\mathcal{V}(t)$. It evaluates the sustainability of the swarm's movement and service strategy.
  \item \textbf{System Capacity ($R_{\text{sum}}$):} The aggregated system throughput as defined in Section \ref{subsubsec:throughput_sum}. This metric reflects the swarm's capability to mitigate co-channel interference and manage the increased load-per-agent following a node failure in dense environments.
\end{itemize}

\begin{figure*}[!t]
    \centering
    \begin{subfigure}[c]{0.335\textwidth}      
      \centering
      \includegraphics[width=\textwidth]{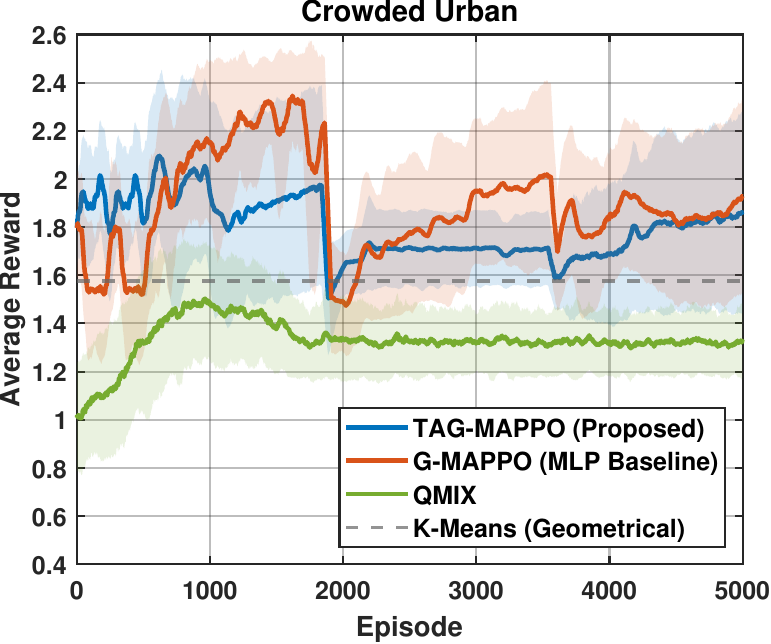}
      \caption{Reward: Crowded Urban}
      \label{fig:convergence_qos:reward_urban}
    \end{subfigure}
    \begin{subfigure}[c]{0.32\textwidth}      
      \centering
      \includegraphics[width=\textwidth]{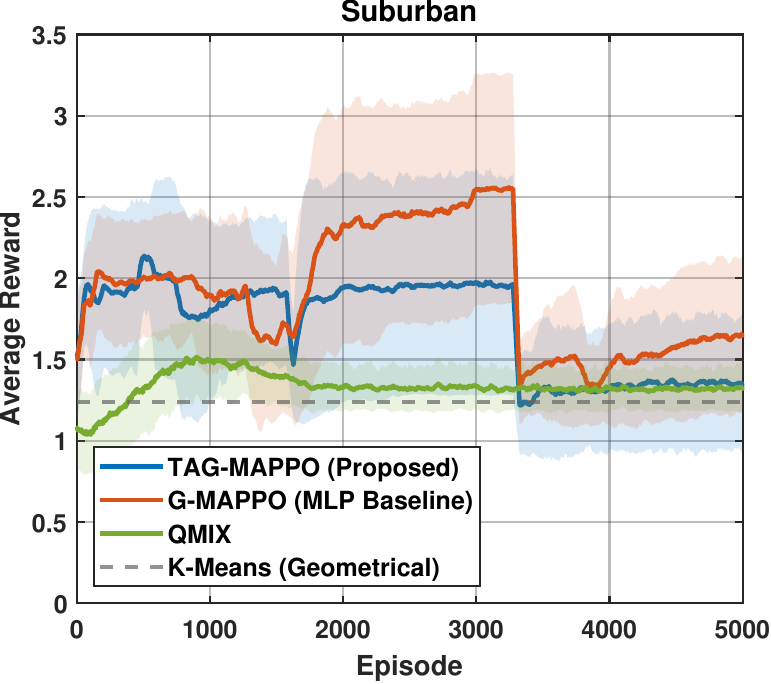}
      \caption{Reward: Suburban}
      \label{fig:convergence_qos:reward_suburban}
    \end{subfigure}
    \hfill
    \begin{subfigure}[c]{0.32\textwidth}
      \centering
      \includegraphics[width=\textwidth]{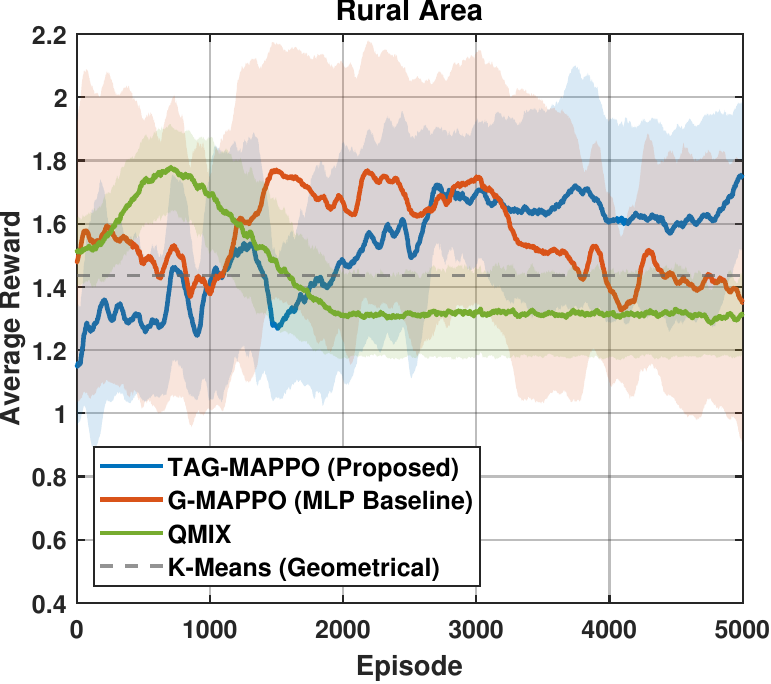}
      \caption{Reward: Rural Area}
      \label{fig:convergence_qos:reward_rural}
    \end{subfigure}\\
    \begin{subfigure}[c]{0.32\textwidth}
      \centering
      \includegraphics[width=\textwidth]{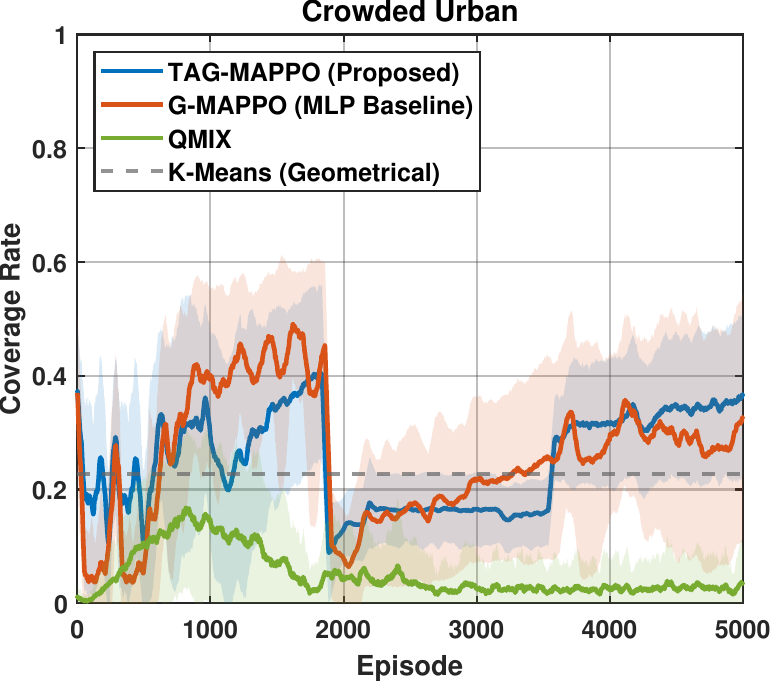}
      \caption{Coverage: Crowded Urban}
      \label{fig:convergence_qos:cov_urban}
    \end{subfigure}
    \hfill
    \begin{subfigure}[c]{0.335\textwidth}
      \centering
      \includegraphics[width=\textwidth]{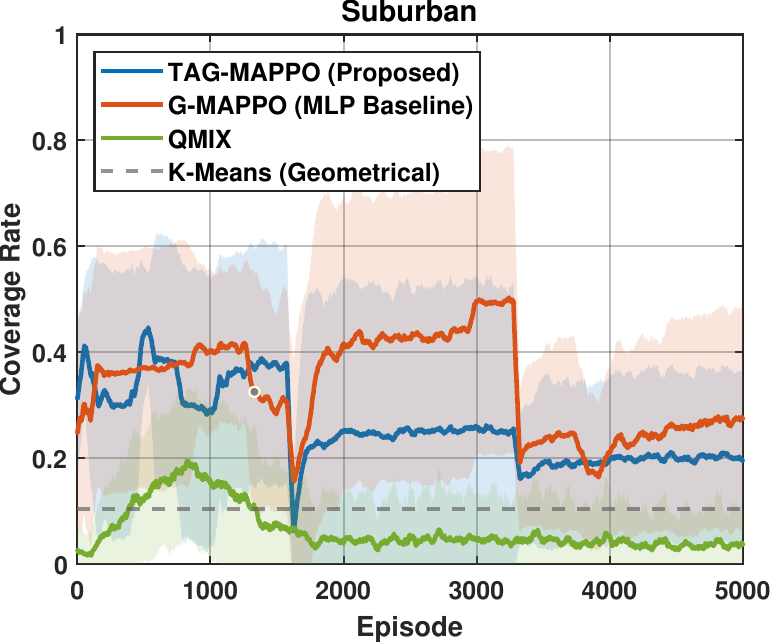}
      \caption{Coverage: Suburban}
      \label{fig:convergence_qos:cov_suburban}
    \end{subfigure}
    \hfill
    \begin{subfigure}[c]{0.32\textwidth}
      \centering
      \includegraphics[width=\textwidth]{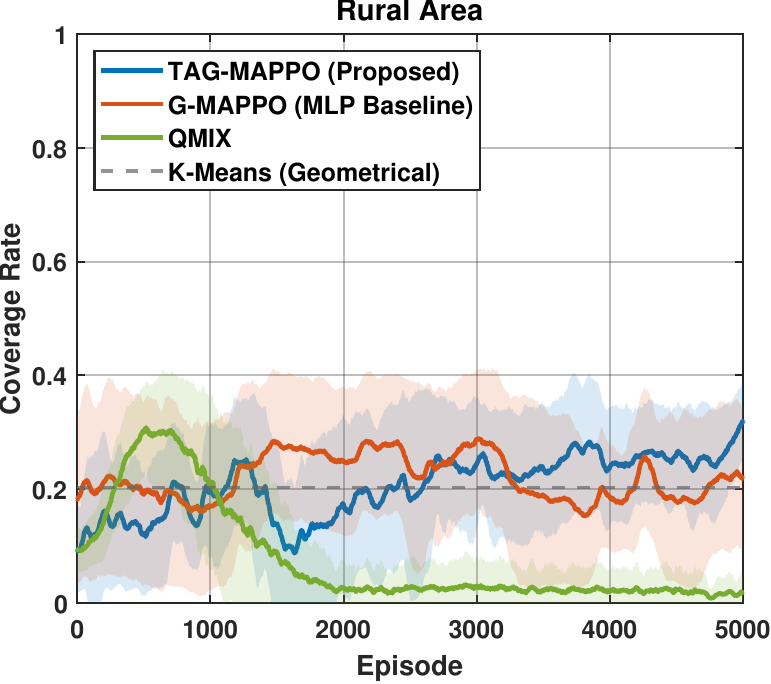}
      \caption{Coverage: Rural Area}
      \label{fig:convergence_qos:cov_rural}
    \end{subfigure}
    \caption{Training convergence analysis across heterogeneous scenarios. The top row, including Figs. \subref{fig:convergence_qos:reward_urban}--\subref{fig:convergence_qos:reward_rural}, illustrates the Average Episode Reward, while the bottom row, Figs. \subref{fig:convergence_qos:cov_urban}--\subref{fig:convergence_qos:cov_rural}, depicts the evolution of the Coverage Ratio ($C_{\text{cov}}$). Results are averaged over 5 independent runs, with shaded areas representing the 95\% confidence interval, demonstrating the framework's stability under stochastic mobility.}
    \label{fig:convergence_qos}
\end{figure*}

\subsection{Training Convergence and QoS Optimization}
\label{subsec:convergence_qos}
The training trajectories, reflecting learning efficacy and multi-objective optimization across three distinct mobility phases, are illustrated in Fig.~\ref{fig:convergence_qos}. To evaluate the framework's robustness, the 5,000-episode training process is divided into: 1) a standard mobility phase ($0 \le ep < 1600$), 2) a dynamic merging phase ($1600 \le ep < 3300$), and 3) a high-speed evacuation phase ($3300 \le ep \le 5000$).

\subsubsection{Learning Efficacy and Convergence Rate} 
As depicted in the top row of Fig.~\ref{fig:convergence_qos}, the \textit{Average Episode Reward} exhibits distinct step-like transitions corresponding to the shifts in user mobility patterns. In the Crowded Urban and Suburban scenarios shown in Fig.~\ref{fig:convergence_qos:reward_urban} and Fig.~\ref{fig:convergence_qos:reward_suburban}, the reward abruptly oscillates at $ep=1600$ as users converge toward the map center. TAG-MAPPO demonstrates superior adaptability, rapidly recalibrating its policy to exploit the increased user density for higher signal-to-interference-plus-noise ratio (SINR).

At $ep=3300$, the transition to the high-speed evacuation mode ($15\text{ m/s}$) triggers a reward drop across all models. However, TAG-MAPPO exhibits a faster recovery slope compared to the G-MAPPO (MLP baseline) as illustrated in Fig.~\ref{fig:convergence_qos:reward_rural}. This advantage stems from the topology-aware attention mechanism, which allows agents to maintain spatial correlation even when user clusters dissipate rapidly. Notably, the QMIX baseline fails to converge in these highly dynamic regimes, underscoring the necessity of policy-based coordination in 3D aerial networks.

\subsubsection{Resilience and Coverage Optimization}
The evolution of the Coverage Ratio $C_{\text{cov}}$ in the bottom row of Fig.~\ref{fig:convergence_qos} provides a physical interpretation of the system's resilience. Despite the inclusion of random node failures during training, TAG-MAPPO maintains a stable $C_{\text{cov}}$ around $0.35$--$0.45$ in urban environments. A significant observation is the reduced variance, represented by the shaded area, of TAG-MAPPO compared to the MLP baseline during the high-speed phase where $ep > 3300$. This indicates that graph-based feature aggregation is inherently more robust to the distribution shifts caused by rapid topological changes.

In Rural Areas, as illustrated in Fig.~\ref{fig:convergence_qos:reward_rural} and Fig.~\ref{fig:convergence_qos:cov_rural}, where user sparsity poses a greater challenge for collaborative tracking, TAG-MAPPO maintains a consistent lead. While G-MAPPO exhibits significant performance collapse in the final stage, TAG-MAPPO effectively prevents coverage voids through its \textit{Residual Ego-State Fusion}, ensuring that agents retain local mission awareness even when neighbor information is sparse. These findings confirm that the proposed framework provides the necessary intelligence for 3D AGINs to maintain Quality of Service (QoS) amidst extreme mobility and unpredictable hardware failures.

\begin{figure*}[!t]
    \centering
    \begin{subfigure}[c]{0.325\textwidth}      
      \centering
      \includegraphics[width=\textwidth]{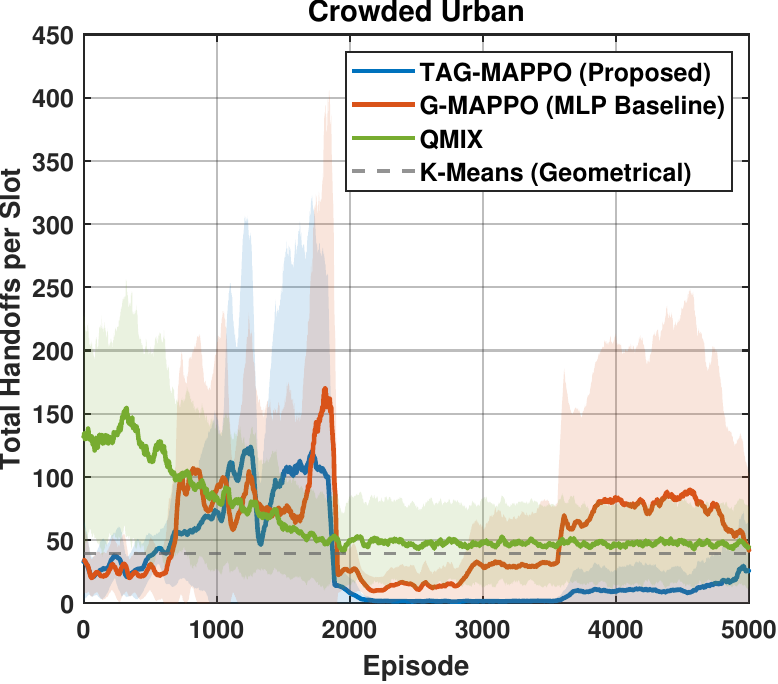}
      \caption{Handoffs: Crowded Urban}
      \label{fig:stability_ee:handoff_urban}
    \end{subfigure}
    \hfill
    \begin{subfigure}[c]{0.325\textwidth}      
      \centering
      \includegraphics[width=\textwidth]{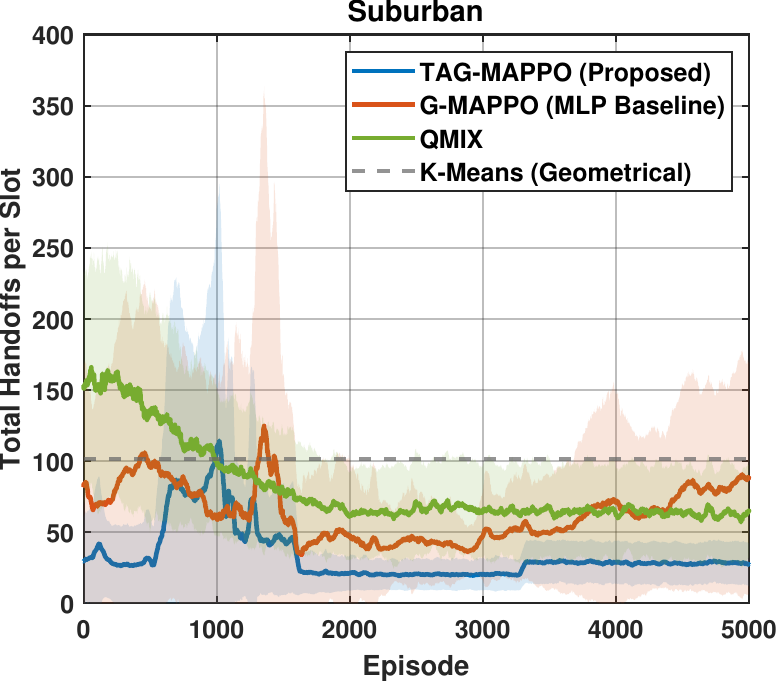}
      \caption{Handoffs: Suburban}
      \label{fig:stability_ee:handoff_suburban}
    \end{subfigure}
    \hfill
    \begin{subfigure}[c]{0.325\textwidth}
      \centering
      \includegraphics[width=\textwidth]{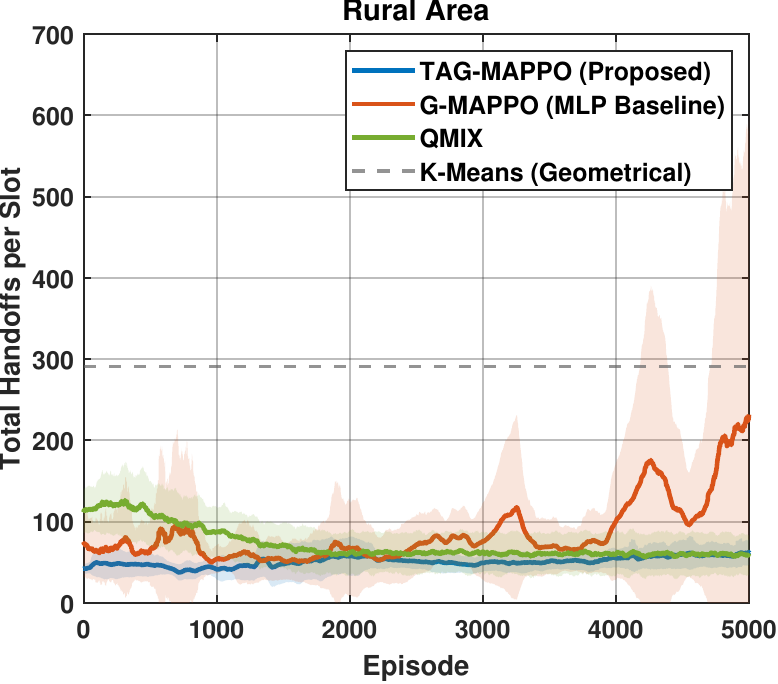}
      \caption{Handoffs: Rural Area}
      \label{fig:stability_ee:handoff_rural}
    \end{subfigure}\\
    \begin{subfigure}[c]{0.325\textwidth}
      \centering
      \includegraphics[width=\textwidth]{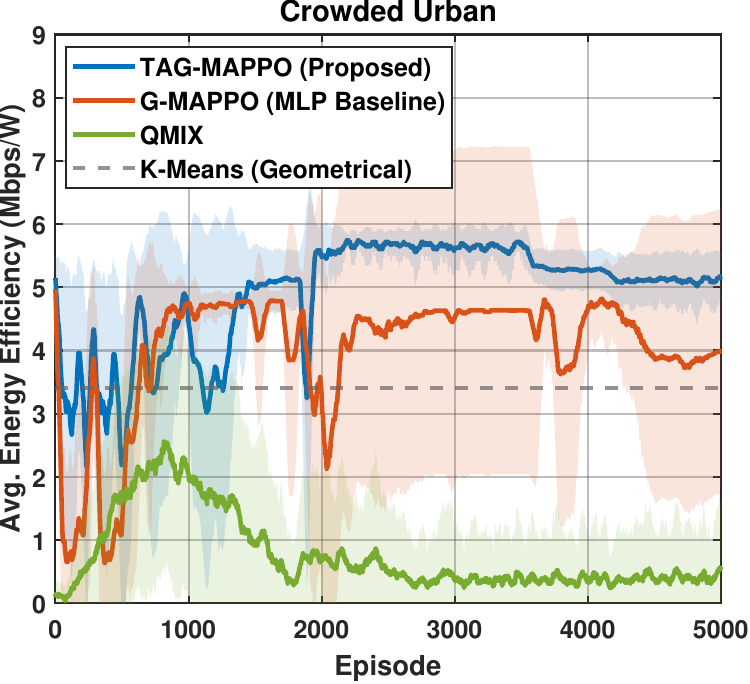}
      \caption{EE: Crowded Urban}
      \label{fig:stability_ee:EE_urban}
    \end{subfigure}
    \hfill
    \begin{subfigure}[c]{0.325\textwidth}
      \centering
      \includegraphics[width=\textwidth]{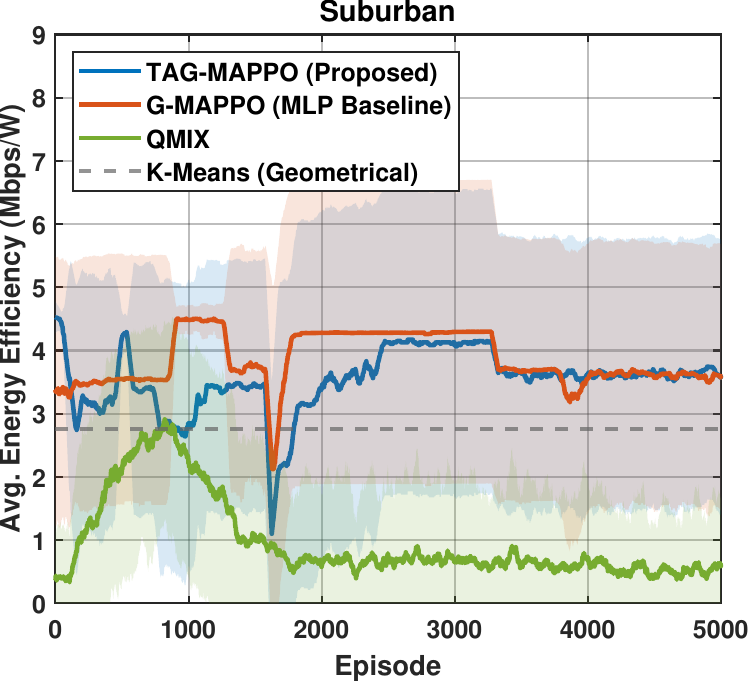}
      \caption{EE: Suburban}
      \label{fig:stability_ee:EE_suburban}
    \end{subfigure}
    \hfill
    \begin{subfigure}[c]{0.325\textwidth}
      \centering
      \includegraphics[width=\textwidth]{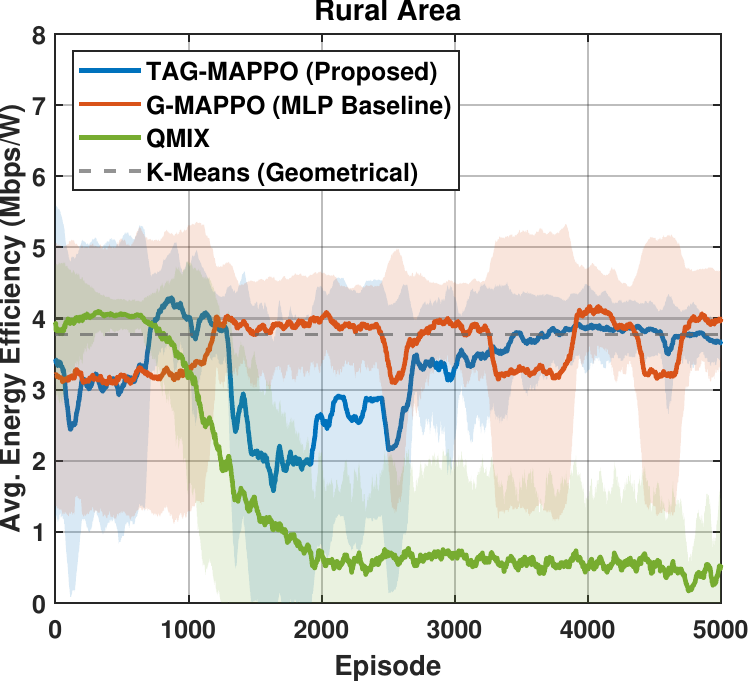}
      \caption{EE: Rural Area}
      \label{fig:stability_ee:EE_rural}
    \end{subfigure}
    \caption{Stability and Energy Efficiency Analysis: The top row, including Figs. \subref{fig:stability_ee:handoff_urban}--\subref{fig:stability_ee:handoff_rural}, illustrates the Total Handoffs ($C_{\text{ho}}$) across different scenarios, while the bottom row, Figs. \subref{fig:stability_ee:EE_urban}--\subref{fig:stability_ee:EE_rural}, depicts the Energy Efficiency ($E_{\text{ee}}$). Results are averaged over 5 independent runs, with shaded areas representing the 95\% confidence interval, demonstrating the framework's superior stability and resource efficiency under stochastic mobility.}
    \label{fig:stability_ee}
\end{figure*}

\subsection{Stability and Resource Efficiency Analysis}
\label{subsec:stability_ee}
The operational sustainability and signaling stability of the TAG-MAPPO framework are evaluated through the \textit{Total Handoffs} $C_{\text{ho}}$ and \textit{Energy Efficiency} $E_{\text{ee}}$ as illustrated in Fig.~\ref{fig:stability_ee}.

\subsubsection{Topological Stability and Handoff Suppression}
The top row of Fig.~\ref{fig:stability_ee} illustrates the framework's capability to suppress excessive signaling overhead amidst environmental transitions. As evidenced in the Crowded Urban and Suburban scenarios shown in Fig.~\ref{fig:stability_ee:handoff_urban} and Fig.~\ref{fig:stability_ee:handoff_suburban}, TAG-MAPPO maintains a remarkably low and stable handoff frequency, consistently staying below 50 events per slot.

A critical observation occurs at the $ep=3300$ transition to high-speed evacuation where the velocity is $15\text{ m/s}$. At this junction, the G-MAPPO baseline suffers from a handoff explosion that peaks near 200 in Rural Areas, as depicted in Fig.~\ref{fig:stability_ee:handoff_rural}. In contrast, TAG-MAPPO remains suppressed throughout the phase. This disparity suggests that the topology-aware attention mechanism allows agents to distinguish between necessary proactive re-associations and erratic ping-pong handoffs caused by rapid user dispersion. Furthermore, the QMIX and MLP baselines exhibit high variance, represented by the shaded areas in Fig.~\ref{fig:stability_ee}, indicating that traditional architectures struggle to maintain connection stability when the underlying graph topology undergoes rapid expansion.

\subsubsection{Resource Utilization and Energy Efficiency}
The bottom row of Fig.~\ref{fig:stability_ee} evaluates the system's throughput-per-Watt performance. In the Crowded Urban scenario shown in Fig.~\ref{fig:stability_ee:EE_urban}, TAG-MAPPO achieves a steady-state efficiency of approximately 5 Mbps/W, which significantly outperforms the K-Means geometrical baseline of 3.4 Mbps/W. This gain is maintained even during the merging phase where $1600 \le ep < 3300$, a period when high user density typically complicates interference management.

The robustness of the proposed framework is further highlighted in the high-speed regime for $ep > 3300$. As shown across all scenarios in Fig.~\ref{fig:stability_ee:EE_urban} through Fig.~\ref{fig:stability_ee:EE_rural}, TAG-MAPPO maintains a tighter confidence interval and higher asymptotic $E_{\text{ee}}$ compared to its competitors. By integrating graph-based features, agents can identify energy-optimal hovering positions that minimize high-speed propulsion power $P_k^{\text{fly}}$ while maximizing communication utility. This consistency in resource efficiency is vital for mission-critical 6G services, as it ensures both battery longevity and control-plane stability in volatile aerial environments.

\begin{figure*}[!t]
    \centering
    \begin{subfigure}[c]{0.325\textwidth}      
      \centering
      \includegraphics[width=\textwidth]{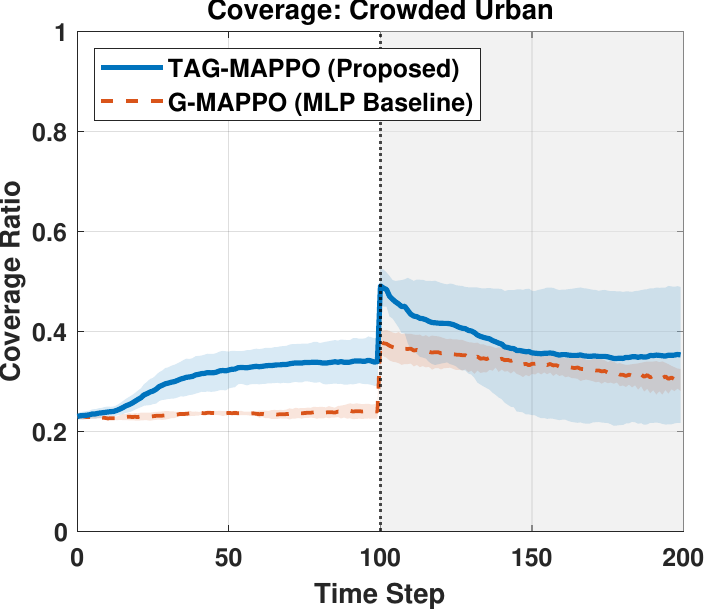}
      \caption{Coverage: Crowded Urban}
      \label{fig:resilience_plots:coverage_urban}
    \end{subfigure}
    \hfill
    \begin{subfigure}[c]{0.325\textwidth}      
      \centering
      \includegraphics[width=\textwidth]{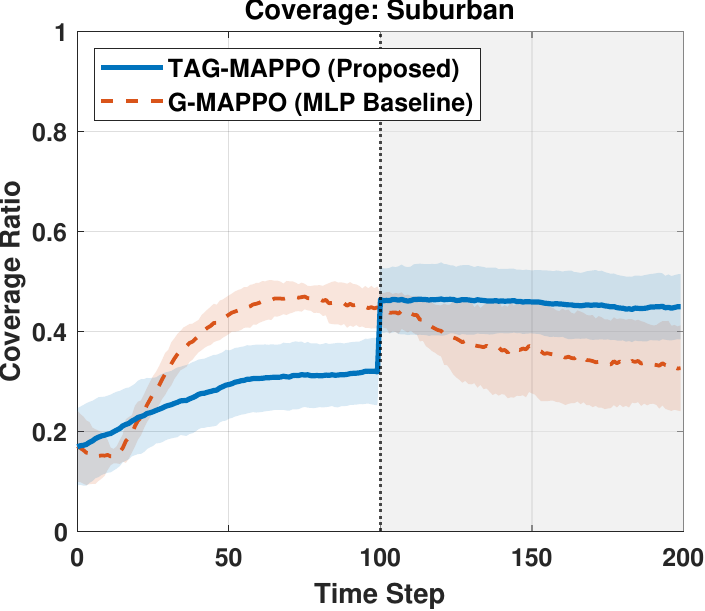}
      \caption{Coverage: Suburban}
      \label{fig:resilience_plots:coverage_suburban}
    \end{subfigure}
    \hfill
    \begin{subfigure}[c]{0.325\textwidth}
      \centering
      \includegraphics[width=\textwidth]{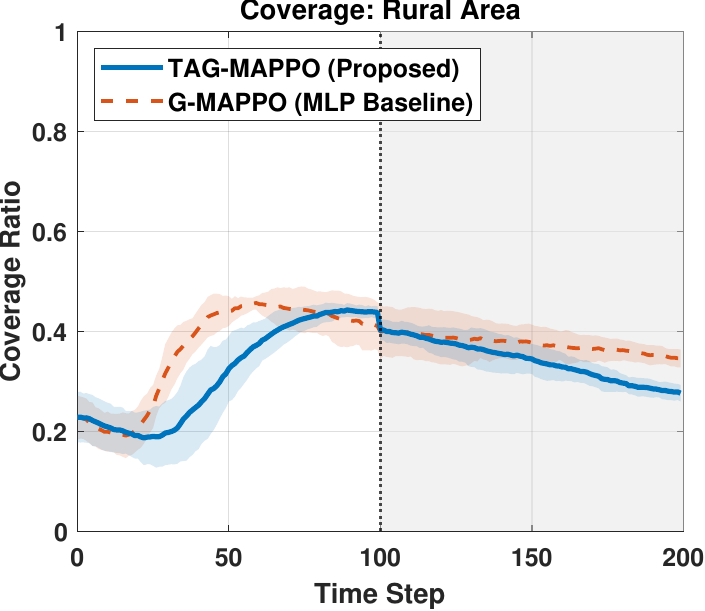}
      \caption{Coverage: Rural Area}
      \label{fig:resilience_plots:coverage_rural}
    \end{subfigure}\\
    \begin{subfigure}[c]{0.325\textwidth}
      \centering
      \includegraphics[width=\textwidth]{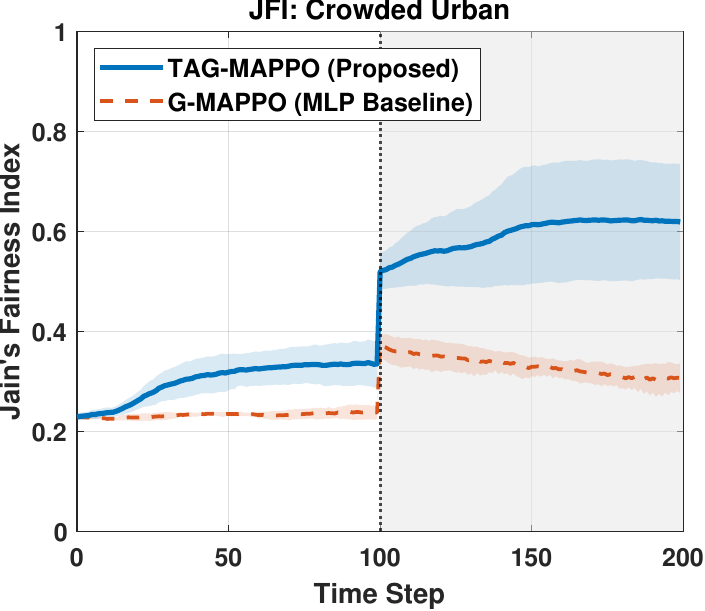}
      \caption{JFI: Crowded Urban}
      \label{fig:resilience_plots:JFI_urban}
    \end{subfigure}
    \hfill
    \begin{subfigure}[c]{0.325\textwidth}
      \centering
      \includegraphics[width=\textwidth]{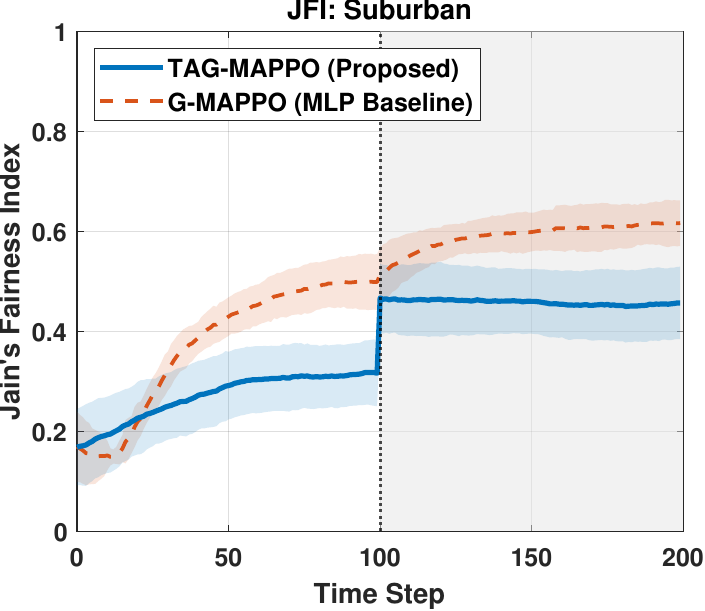}
      \caption{JFI: Suburban}
      \label{fig:resilience_plots:JFI_suburban}
    \end{subfigure}
    \hfill
    \begin{subfigure}[c]{0.325\textwidth}
      \centering
      \includegraphics[width=\textwidth]{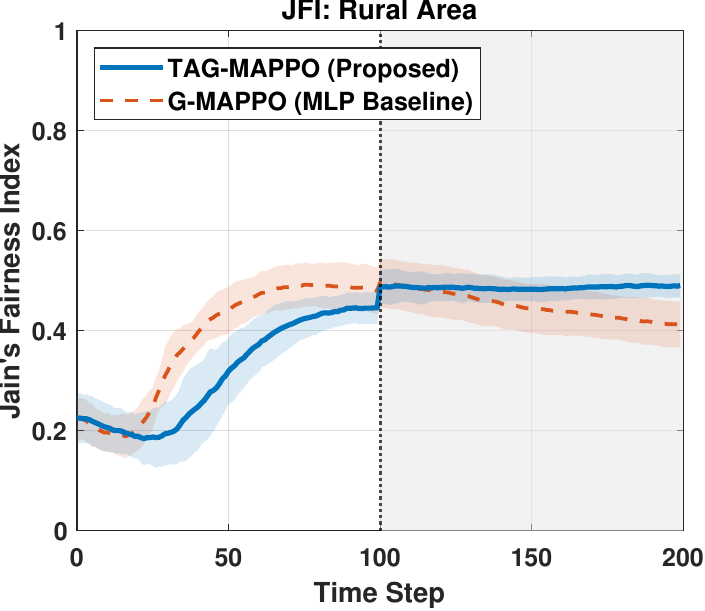}
      \caption{JFI: Rural Area}
      \label{fig:resilience_plots:JFI_rural}
    \end{subfigure}
    \caption{System Resilience Analysis: The top row, including Figs. \subref{fig:resilience_plots:coverage_urban}--\subref{fig:resilience_plots:coverage_rural}, illustrates the Coverage Ratio ($C_{\text{cov}}$) following a random node failure at $t=100$, while the bottom row, Figs. \subref{fig:resilience_plots:JFI_urban}--\subref{fig:resilience_plots:JFI_rural}, depicts the Jain's Fairness Index (JFI) over the same period. Results are averaged over 5 runs with shaded areas representing the standard deviation.}
    \label{fig:resilience_plots}
\end{figure*}

\subsection{System Resilience under Catastrophic Node Failure}
\label{subsec:resilience_analysis}
To evaluate the survivability of the proposed framework, we analyze the system performance following a random node failure at $t=100$. A fundamental advantage of TAG-MAPPO is its inherent capability to handle dynamic changes in the agent population. Unlike conventional MARL architectures such as QMIX, which rely on a centralized mixing network with fixed input dimensions, our framework leverages the permutation-invariant nature of graph-based feature aggregation.

In a QMIX-based system, the loss of a node at $t=100$ leads to a structural mismatch in the mixing layer, which makes it unsuitable for resilience tests without manual reconfiguration. In contrast, TAG-MAPPO processes the network as a graph of relational entities, allowing the surviving UAVs to maintain stable coordination even when a vertex is abruptly removed. This flexibility and the subsequent recovery process are visualized in Fig.~\ref{fig:resilience_plots}.

\subsubsection{Dynamic Recovery and Coverage Self-healing}
As illustrated in the top row of Fig.~\ref{fig:resilience_plots}, TAG-MAPPO exhibits an immediate and robust response to the catastrophic event at $t=100$. In the Crowded Urban and Suburban scenarios shown in Fig.~\ref{fig:resilience_plots:coverage_urban} and Fig.~\ref{fig:resilience_plots:coverage_suburban}, the Coverage Ratio $C_{\text{cov}}$ does not suffer from a persistent collapse. Instead, it maintains a stable level that significantly outperforms the G-MAPPO baseline.

This superior resilience is achieved through the integration of the \textit{Random Observation Shuffling} (ROS) mechanism during the training phase. By stochastically permuting the sequence of neighbor observations, ROS prevents the agents from developing a coordinate-index dependency where the policy implicitly relies on the fixed ordering of neighbor information. Consequently, when a node failure occurs, the surviving agents do not perceive the missing data as a broken input but rather as a valid topological variation.

This structural generalization, reinforced by exposure to stochastic failures during training, allows the swarm to internalize a robust coordination logic that adapts to unpredictable topological ruptures. Upon the failure of a node, the remaining agents immediately re-evaluate the transformed topology through the attention mechanism to fill the service gap, ensuring that the aggregate $C_{\text{cov}}$ remains competitive despite the 25 percent reduction in available hardware resources.

\subsubsection{Structural Reorganization and Performance Gain}
Beyond mere coverage restoration, the bottom row of Fig.~\ref{fig:resilience_plots} highlights a remarkable phenomenon in Jain's Fairness Index $\mathcal{J}_{\text{r}}$. Following the node failure in the Crowded Urban scenario depicted in Fig.~\ref{fig:resilience_plots:JFI_urban}, the $\mathcal{J}_{\text{r}}$ of TAG-MAPPO actually surpasses its pre-failure level, rising from 0.35 to 0.6.

This counter-intuitive performance gain suggests that the proposed framework achieves a more efficient spatial equilibrium after the failure. In high-density environments, the removal of a redundant or interfering node allows the surviving agents to eliminate service overlaps and coordination conflicts. The ROS mechanism plays a pivotal role here by ensuring that the feature aggregation logic treats the neighboring swarm as a flexible set rather than a fixed vector. By autonomously re-optimizing their positions, the remaining swarm mitigates co-channel interference that was sub-optimally handled in the original four-agent configuration. This achievement of a more efficient spatial equilibrium is particularly significant as it demonstrates the model's ability to identify and resolve service redundancies in real time.

In contrast, the results for the Rural Area shown in Fig.~\ref{fig:resilience_plots:JFI_rural} show that the MLP baseline fails to reorganize, with its $\mathcal{J}_{\text{r}}$ collapsing towards 0.25. This failure indicates that traditional architectures struggle to maintain connection stability when the underlying graph topology undergoes rapid contraction. TAG-MAPPO, however, maintains a significantly higher fairness index with a tighter confidence interval. These results confirm that graph-based reasoning enables UAVs to maintain service survivability even in hostile operational conditions. The integration of residual ego-state fusion ensures that each agent retains sufficient self-awareness to track user clusters independently, satisfying the critical requirements for 6G mission-critical reliability during hardware malfunctions.

\section{Conclusion}
\label{sec:conclusion}
This paper presented TAG-MAPPO, a topology-aware reinforcement learning framework designed to optimize service quality and operational stability in 3D Aerial-Ground Integrated Networks. By integrating graph-based feature aggregation with a residual ego-state fusion mechanism, the framework successfully deciphers complex spatial dependencies among UAV agents in dynamic environments. Our approach addresses fundamental challenges of signaling overhead and resource inefficiency that hinder autonomous aerial base station deployment in 6G scenarios. Experimental results across three geographical profiles demonstrate that TAG-MAPPO outperforms conventional MLP-based MAPPO and QMIX baselines. By implementing the Random Observation Shuffling (ROS) mechanism, the framework reduces redundant signaling by up to 50 percent in sparse rural topologies while maintaining superior energy efficiency and stability during high-speed mobility phases.

The system resilience tests highlight the structural generalization capability of TAG-MAPPO, where the surviving swarm autonomously reconfigures its topology to restore coverage immediately following catastrophic node failures. In dense urban scenarios, the framework achieves a higher fairness index post-failure compared to its original four-UAV configuration, illustrating the attention mechanism's ability to resolve service overlaps once interference sources are removed. These findings confirm that topological intelligence is essential for robust and scalable coordination in 6G mission-critical missions. Future work will extend this framework to multi-tier architectures involving satellite-UAV coordination. Additionally, integrating secure communication protocols within the topology-aware learning process remains a valuable direction to ensure the integrity of the Internet of UAVs against adversarial interference.




\bibliographystyle{IEEEtran}
\bibliography{reference}
\ifCLASSOPTIONcaptionsoff  \newpage \fi 

\begin{IEEEbiography}[{\includegraphics[width=1in,height=1.25in,clip,keepaspectratio]{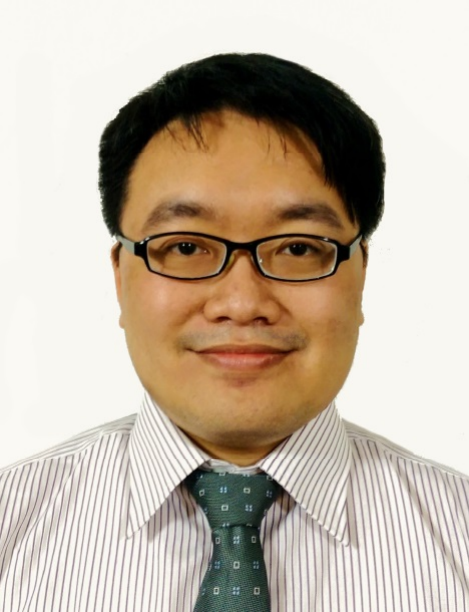}}]{Chuan-Chi Lai}
    (Member, IEEE) received the Ph.D. degree in Computer Science and Information Engineering from the National Taipei University of Technology, Taiwan, in 2017. He held research and faculty positions at National Chiao Tung University and Feng Chia University prior to his current role. Since 2024, he has been an Assistant Professor with the Department of Communications Engineering, National Chung Cheng University, Chiayi, Taiwan. His research interests include mobile edge computing, UAV networks, and AI for wireless communications. Dr. Lai was a recipient of the Postdoctoral Researcher Academic Research Award from the NSTC, Taiwan, in 2019, and Best Paper Awards at WOCC (2018, 2021) and ICUFN (2015).
\end{IEEEbiography}

\end{document}